\documentclass[a4paper,12pt]{article}
\usepackage[centertags]{amsmath}
\usepackage{amstext}
\usepackage{graphicx}
\usepackage{epsfig,verbatim}
\usepackage{caption}
\usepackage{subcaption}
\usepackage{cite}

\usepackage[english]{babel}
\usepackage{amssymb,amsfonts}
\numberwithin{equation}{section}
\newcommand{\be}{\begin{equation}}
\newcommand{\ee}{\end{equation}}
\newcommand{\bea}{\begin{eqnarray}}
\newcommand{\eea}{\end{eqnarray}}

\newcommand{\p}{\partial}
\newcommand{\bc}{\begin{center}}
\newcommand{\ec}{\end{center}}
\newcommand{\ie}{{\it i.e.~}}

\newcommand{\half}{\frac{1}{2}}

\renewcommand{\Re}{{\rm Re}\,}
\renewcommand{\Im}{{\rm Im}\,}

\newcommand{\tr}{{\rm tr}}
\newcommand{\mb}{\mathbf}
\def\imath{i}

\title{\hfill\parbox{4cm}{\normalsize TAUP-2292-15 \\ FPAUO-15/03\\
}\\
\vspace{2cm}
\bf New knotted solutions of Maxwell's equations}

\author{\small Carlos Hoyos$^1$, Nilanjan Sircar$^{2}$ and Jacob Sonnenschein$^{2}$\\
\begin{tabular}{l}
\footnotesize $^1$Department of Physics, Universidad de Oviedo,  Avda. Calvo Sotelo 18, 33007, Oviedo, Spain.\\
\footnotesize $^2$Raymond and Beverly Sackler School of Physics and Astronomy, Tel-Aviv University, Tel-Aviv 69978, Israel.
\end{tabular}\\
\small{carlos.hoyos.badajoz@gmail.com, nilanjan.tifr@gmail.com, and cobi@post.tau.ac.il}
          }
\date{}
\begin{document}
\maketitle
\begin{abstract}
In this note we have further developed the study of topologically non-trivial solutions of vacuum electrodynamics. 
We have discovered a novel method of generating such solutions  by applying conformal transformations with 
complex parameters  on known solutions expressed in terms of Bateman's variables.
This has enabled us  to get a wide class of  solutions  from the basic configuration like constant electromagnetic fields 
and plane-waves. We have introduced a covariant formulation of the Bateman's construction and 
discussed the conserved charges associated with the conformal group as well as a set of four types of conserved helicities. 
We have also given a formulation in terms of quaternions.
This led to a simple map between the electromagnetic knotted and linked solutions into flat connections of $SU(2)$ 
gauge theory. We have computed the corresponding Chern-Simons charge in a class of solutions and it takes integer values. 
\end{abstract}

\newpage
\tableofcontents
\section{Introduction}

It is ``common wisdom" that  topological non-trivial  solutions  in field theory associate only with non-linear equations of motion. 
That is the case for instance for two dimensional solitons, magnetic monopoles and four dimensional Yang-Mills' instantons. 
In fact it turns out that this statement is false and  there are also non-trivial configurations which solve linear equations 
of motion.  In particular there are known  solutions of free  Maxwell's equations in flat space-time that admit non-trivial 
topology, the Hopfion solution constructed in \cite{Ranada:1989wc}. Since then there have been considerable study of these solutions ~\cite{Ranada:1990,Ranada:1992hw,Ranada:1995,Ranada:1997},
which was recently revived in ~\cite{IrvineBouwmeester2008,Irvine2010,ArrayasTrueba,Arrayas:2011ci,Arrayas:2011ia}. These solutions were
studied as null electromagnetic solutions in Bateman's ~\cite{Bateman} construction in
~\cite{Besieris,PhysRevLett.111.150404}. 
The generalization and classification of analog solutions in higher spin fields was studied in
~\cite{Dalhuisen:2012zz,Swearngin:2013sks,Thompson:2014pta,Thompson:2014owa}. Linked and knotted non-null solutions of electrodynamics were studied in~\cite{Arrayas:2015}. A collection of comprehensive notes and references of such solutions in electrodynamics and other areas of physics can be found in~\cite{Hopfion.com}.

Since the original determination by Maxwell  of the equations of motion of Electrodynamics, many methods have been proposed to determine the  evolution in space and time of the electric and magnetic fields. 
The method we will use in this note is the so called {\bf Bateman's construction} ~\cite{Bateman} 
which is based on grouping the electric and magnetic fields into a complex vector field (Riemann-Silberstein vector~\cite{bialynicki1996photon, bialynicki2013role}), 
and expressing it in terms of two complex functions $\alpha$ and $\beta$ as follows,
\begin{equation}
\mb{F}=\mb{E} + \imath \mb{B} \qquad \mb{F}=\nabla\alpha \times \nabla\beta.
\end{equation}
For this ansatz of the fields the solutions of the equations of motion are necessarily null namely
\begin{equation}
\mb{F}^2=0\ \ \Rightarrow \mb{E}^2-\mb{B}^2=0,\ \ \mb{E}\cdot\mb{B}=0.
\end{equation} 
Certain solutions of the equations of motion have a very cumbersome  form when expressed directly in terms of 
$\mb{E}$ and $\mb{B}$ but are very compact  in terms of $\alpha$ and $\beta$. An example for that is the 
 remarkable solution  which admits a non-trivial linking between the electric and magnetic fields, 
 which  takes the form\cite{PhysRevLett.111.150404},
\be\label{basicsol}
\alpha = \frac{A-1+ \imath z}{A+  \imath t} \ , \qquad
\beta = \frac{(x-\imath y) }{A+ \imath t}.
\ee
where $A=\half(x^2+y^2+z^2-t^2+1)$. The solution has an implicit dependence on a scale that we have set to one. We can make the dependence explicit by rescaling the coordinates $(t,x,y,z)\longrightarrow \lambda(t,x,y,z)$. 
Note also that the solution is valid for any conformally flat metric, since Maxwell's equations do not change. This space and time dependent solution has finite energy, momentum, angular momentum, etc. 
 The topological nature of this solution manifest itself in the fact that it carries non-trivial ``Chern-Simons (CS)  helicity charges".
 Exactly the same charges are also characterizing the so called {\bf  Hopfion} solution of Maxwell's equation
 \cite{Ranada:1989wc},\cite{Ranada:1990} which can be derived from the basic Hopf map $S^3\rightarrow  S^2$. 
 It was shown in \cite{Besieris, PhysRevLett.111.150404} that in fact there is an infinitely large family of solutions that can be 
 derived from the basic one by mapping $\alpha\rightarrow f(\alpha, \beta)$ and $\beta \rightarrow g(\alpha,\beta)$,
 where $f$ and $g$ are two arbitrary holomorphic functions.
One may wonder if there are other ways to easily generate new  families of solutions, in this note we show that the answer is 
in the affirmative. 
We will prove that any transformation associated with the full conformal group $SO(2,4)$ but with complex rather than real parameters of a given solution is also a solution of Maxwell's equation. 
Obviously such transformations are not symmetries of the Maxwell's action however they do yield new solutions. 
Implementing this idea we found out that applying  a temporal special conformal transformation with an imaginary parameter on a constant electric and magnetic perpendicular fields produces the topologically non-trivial solution of (\ref{basicsol}).
We further generated new solutions by applying such transformations on plane waves and on the Hopfion (\ref{basicsol}) itself. One remarkable property of these transformations is that they map solutions of infinite energy and Noether charges to solutions with finite energy and charges.

We have  discovered another interesting property of the knotted solutions of Maxwell's equation. They can be mapped into $SU(2)$ flat gauge connections. Defining a quaternionic valued function 
\be
\mb{q} = \frac{1}{\sqrt{|\alpha|^2+|\beta|^2}} (\alpha + \beta j)
\ee
We can map the unit quaternions to elements of $SU(2)$ using a representation in terms of Pauli matrices.
Then, we write down a flat gauge connection as $\mb{Q}_{\mu} =\mb{q}(\partial_{\mu} \mb{q}^*) $.  We determine the winding number  from the calculation of the non-Abelian Chern-Simons.

The paper is organized as follows.  In the next section we review Bateman's construction of Maxwell's equations. We write down  the variables, the constraint equations and the method to deduce from them the electric and magnetic fields. 
The conserved charges associated with the conformal symmetry group  that characterize the solutions are written down. 
We then describe, using Bateman's construction variables,  three  particular solutions: 
(i)  Constant perpendicular electric and magnetic fields,
(ii)   The plane wave solution,
(iii)  The Hopfion solution. The latter is the basic prototype solution that admits non-trivial topology.  In section 3 we describe the new method for  generating  novel non-trivial solutions. The method  utilizes  transformations of the full conformal group with complex rather than real parameters. We first prove that indeed the transformed configurations are solutions of the equations of motion. We then show in section 3.2 how one can get the  topologically non-trivial $(p,q)$ solutions from a constant electric and magnetic fields.  Next we get knotted solutions from plane waves and in subsection 3.4 we apply the complex transformation on  the Hopfion configuration in particular time translations, space translations, rotations, boosts, scalings and special conformal transformations. Section 4 is devoted to additional formal  aspects of Bateman's construction and
knot solutions. In subsection 4.1 we write down a covariant formulation of the construction and then in subsection 4.2 we derive general formulas for the helicities of the solutions.  In section 4.3 we reformulate Bateman's construction using quaternionic fields. We show that this naturally leads to a relation between the knotted Abelian solutions and flat $SU(2)$ non-Abelian gauge connections with non-zero winding number.
Section 5 is devoted to a summary and a list of open questions. Three appendices are added: In  the first  the Noether charges are determined, the second describes a different formulation of the Hopfion solutions and in the third we write down explicit expressions for the components of the electric and magnetic fields.

\section{Bateman's construction}

A procedure to construct topologically non-trivial solutions in Maxwell's equations is to use Bateman's construction 
\cite{Bateman}. In 3+1 dimensions we can group the electric and magnetic fields in a single complex vector field, 
the Riemann-Silberstein vector \cite{bialynicki1996photon, bialynicki2013role}
\begin{equation}
\mb{F}=\mb{E} + \imath \mb{B}.
\end{equation}
In the absence of charged matter, Maxwell's equations for the complex field are
\begin{equation}
\nabla\cdot \mb{F}=0,\ \ i\partial_t \mb{F}=\nabla\times \mb{F}.
\end{equation}
We can satisfy the divergenceless condition with the general ansatz
\begin{equation} \label{absolanz1}
\mb{F}=\nabla\alpha \times \nabla\beta,
\end{equation}
where $\alpha$ and $\beta$ are complex. Then, the dynamical equation becomes
\begin{equation}
i\nabla\times(\partial_t \alpha \nabla\beta-\partial_t \beta \nabla\alpha)=\nabla\times \mb{F}.
\end{equation}
Which will be satisfied if 
\begin{equation}\label{eomab}
i(\partial_t \alpha \nabla\beta-\partial_t \beta \nabla\alpha)=\mb{F}=\nabla\alpha\times \nabla\beta.
\end{equation}
Solutions of this form have null norm
\begin{equation}
\mb{F}^2=i(\partial_t \alpha \nabla\beta-\partial_t \beta \nabla\alpha)\cdot \left(\nabla\alpha\times \nabla\beta\right)=0\ \ \Rightarrow \mb{E}^2-\mb{B}^2=0,\ \ \mb{E}\cdot\mb{B}=0.
\end{equation}
For this reason, they are also known as null solutions~\cite{Besieris,PhysRevLett.111.150404}.

\subsection{Conserved charges}

We will characterize solutions by their conserved charges. In the first place we have the usual conserved charges: energy $E$, momentum $\mb{P}$ and angular momentum $\mb{L}$. They can be written as integrals over space of the following gauge-invariant charge densities
\begin{itemize}
\item Energy density: $\mathcal{E} = \half (\mb{E}^2+\mb{B}^2)$,  
\item Momentum density:  $ \mb{ p} = (\mb{E} \times \mb{B})$,
\item Angular momentum density: $\mb{\ell} =(\mb{p} \times \mb{x})$.
\end{itemize}
Since Maxwell's theory is invariant under conformal transformations, we have additional charge densities associated to the following transformations
\footnote{The charge densities for special conformal transformation given here differ from that in \cite{IrvineBouwmeester2008}, although
their value agree at $t=0$. We provide a derivation of the right charges in Appendix ~\ref{appendixCharges}. We have explicitly checked that they are time-independent.}
\begin{itemize}
\item Boosts: $\mb{b}_v = (\mathcal{E} \mb{x} - \mb{P} t)$,
\item Dilatations:  $d= (\mb{p}\cdot \mb{x}-\mathcal{E} t)$,
\item Temporal special conformal transformations (TSCT): $k^0 = \left((x^2+t^2) \mathcal{E}- 2 t \mb{p}\cdot \mb{x}\right)$,
\item Spatial special conformal transformations (SSCT): $\mb{k} = \left(2 \mb{x} (\mb{p}\cdot \mb{x})-2 t \mathcal{E} \mb{x}-(x^2-t^2) \mb{p} \right)$. 
\end{itemize}
We will denote the conserved charges as $\mb{B}_v$, $D$, $K^0$ and $\mb{K}$ respectively.

For the null solutions we have an additional set of conserved charges that do not have a gauge-invariant density, the electric and magnetic helicities. We define them as the integral over space of the Chern-Simons form of the gauge potentials.

Let us first define the complex vector potential $\mb{V}$, such that $\mb{F}=\nabla\times \mb{V}$. We can decompose it in real and imaginary parts $\mb{V}=\mb{C}+i\mb{A}$, so it is clear that $\mb{A}$ is the usual gauge potential and $\mb{C}$ is the `dual' potential in the sense of electromagnetic duality 
\begin{equation}
\mb{E}=\nabla\times \mb{C},\ \ \mb{B}=\nabla\times \mb{A}.
\end{equation}
From Eq.~(\ref{absolanz1}), we can derive the relation between the gauge potentials and the functions $\alpha$ and $\beta$. Up to gauge transformations
\be\label{CplusiA}
\mb{V}=\mb{C} + \imath \mb{A} = \alpha \nabla \beta .
\ee
Then,
\begin{equation}
\mb{C}=\text{Re}\,(\alpha\nabla\beta),\ \ \mb{A}=\text{Im}(\alpha\nabla\beta).
\end{equation}
In this language, electromagnetic duality is simply the transformation $\alpha\to \imath \alpha$ or $\beta\to \imath\beta$, which clearly is a symmetry of the equations of motion \eqref{eomab}.

The electric and magnetic helicities are defined as
\begin{equation}
h_{ee}=\int d^3x\,\mb{C}\cdot\mb{E},\ \ h_{mm}=\int d^3 x\, \mb{A}\cdot \mb{B}.
\end{equation}
They are gauge invariant when the integral is over all space, since $\nabla\cdot \mb{E}=\nabla\cdot \mb{B}=0$.  In addition, we can define the cross helicities
\begin{equation}
h_{em}=\int d^3x\,\mb{C}\cdot\mb{B},\ \ h_{me}=\int d^3 x\, \mb{A}\cdot \mb{E}.
\end{equation}
They are also gauge invariant when the integral is over all space.

To show that the helicities are conserved for the null solutions is very simple. For the magnetic helicity we have
\begin{eqnarray}
\partial_t h_{mm} &=& \int d^3x\,\left(\partial_t\mb{A}\cdot\mb{B}+\mb{A}\cdot\partial_t\mb{B}\right) \nonumber \\
&=& -\int d^3x\,\left((\mb{E}+\nabla \Phi)\cdot\mb{B}+\mb{A}\cdot(\nabla\times \mb{E})\right)\nonumber \\
&=& \int d^3 x\, \left(\Phi \nabla\cdot\mb{B}-\mb{E}\cdot (\nabla\times \mb{A})\right)=-\int d^3 x\, \mb{E}\cdot \mb{B}=0.
\end{eqnarray}
We have used that $\mb{E}=-\partial_t\mb{A}-\nabla\Phi$ and $\mb{E}\cdot\mb{B}=0$ ($\mb{A}$ and $\phi$ are vector and scalar
electromagnetic potential).  We can repeat the same calculation for the electric helicity $h_{ee}$ using the dual potential. For the cross helicities we can also show that they are conserved. For $h_{me}$
\begin{eqnarray}
\partial_t h_{me} &=& \int d^3x\,\left(\partial_t\mb{A}\cdot\mb{E}+\mb{A}\cdot\partial_t\mb{E}\right) \nonumber\\
&=& \int d^3x\,\left((-\mb{E}-\nabla \Phi)\cdot\mb{E}+\mb{A}\cdot(\nabla\times \mb{B})\right) \nonumber \\
&=& \int d^3 x\, \left(-\mb{E}^2+\Phi \nabla\cdot\mb{E}+\mb{B}\cdot (\nabla\times \mb{A})\right)=-\int d^3 x\, (\mb{E}^2- \mb{B}^2)=0.
\end{eqnarray}
And a similar calculation can be done for $h_{em}$.

In fact, for solutions found using Bateman's construction, the two helicities have the same value and the two cross helicities have opposite sign. One can show it using that the complex potential is orthogonal to the complex field strength
\begin{equation}
\mb{V}\cdot\mb{F}=\alpha\nabla\beta\cdot (\nabla\alpha\times \nabla\beta)=0\ \ \Rightarrow \mb{C}\cdot \mb{E}-\mb{A}\cdot \mb{B}=0,\ \  \mb{C}\cdot \mb{B}+\mb{A}\cdot \mb{E}=0 .
\end{equation}
Therefore, all the solutions will have $h_{ee}=h_{mm}$, $h_{me}=-h_{em}$.

 \subsection{Examples}\label{sec:knownsolution}

There are several kind of known solutions to Maxwell's equations that can be found using Bateman's construction. 
The simplest solution one can consider is that corresponding to constant electric and magnetic field,
\be
\alpha= 2 \imath (t+z)-1 ~~;~~ \beta = 2 ( x- \imath y) \label{abconstantF}.
\ee
The corresponding electric and magnetic field are null and perpendicular to each other,
\be
\mb{E}= (-4,0,0);~~~\mb{B}=(0,4,0),
\ee
This solution has diverging total energy and total helicity.

The next simplest are plane waves. 
An example is\cite{PhysRevLett.111.150404},
\be
\alpha = e^{\imath (z-t)} ~~;~~ \beta = x + \imath y \label{PWalphabeta}.
\ee
Then,
\be
\mb{F}= (\hat x + \imath \hat y) e^{\imath (z-t)}.
\ee
This is a linearly polarized plane wave moving along the $z$ direction. 
These solutions have finite energy density and momentum, but all the charges diverge when integrated over the full space. 
Therefore, we cannot associate well defined helicities to them.

A more interesting kind of solution is the Hopfion, that we describe in detail in the Appendix~\ref{secHopfionI}. The Hopfion was originally found in \cite{Ranada:1989wc}, and the same solution was constructed again later using Bateman's construction in \cite{PhysRevLett.111.150404}.\footnote{This solution (field strength) is the same up to a trivial constant factor
as that described in the Appendix~\ref{secHopfionI}, after the identification $t \to -t$ and $\phi \leftrightarrow \theta$.}
\bea
\alpha &=& \frac{A-1+ \imath z}{A+  \imath t} \label{hopfalpha1}, \\
\beta &=& \frac{(x-\imath y) }{A+ \imath t}\label{hopfbeta1}.
\eea
where $A=\half(x^2+y^2+z^2-t^2+1)$. In this form, the functions $\alpha$ and $\beta$ satisfy the additional condition $|\alpha|^2+|\beta|^2=1$, so for fixed time they can be seen as the inverse of a stereographic map of $S^3$ to $\mathbb{R}^3\cup \{\infty\}$. We use this later to give a general formula for the helicity that is independent of the functional form of $\alpha$ and $\beta$.
We must point out that although this is a nice representation, it is not unique, in other representations 
$|\alpha|^2+|\beta|^2$ may not be a constant~\cite{PhysRevLett.111.150404}. The explicit form of the components of the 
electric and magnetic fields  is given in appendix \ref{appendixExplicitEB} (similar expressions can also be found in the original work \cite{Ranada:1990}.
See also \cite{Thompson:2014owa} for a review).

\begin{figure}
    \centering
    \begin{subfigure}[b]{0.4 \textwidth}
        \centering
        \includegraphics[width=\textwidth]{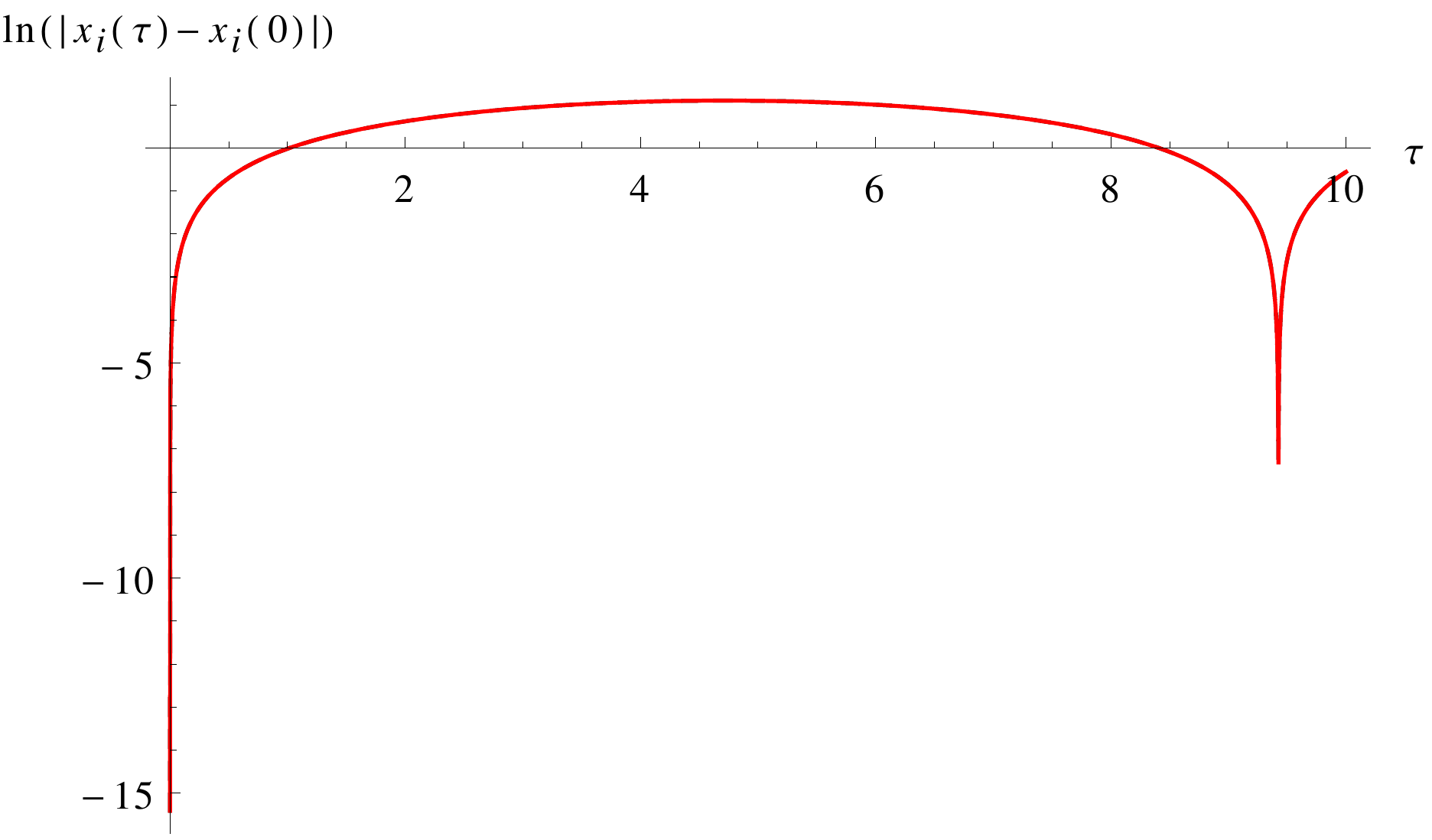}
        \caption{}
        \label{fig:HopfionEBFLdis}
    \end{subfigure}
    \hfill
    \begin{subfigure}[b]{0.4 \textwidth}
        \centering
        \includegraphics[width=\textwidth]{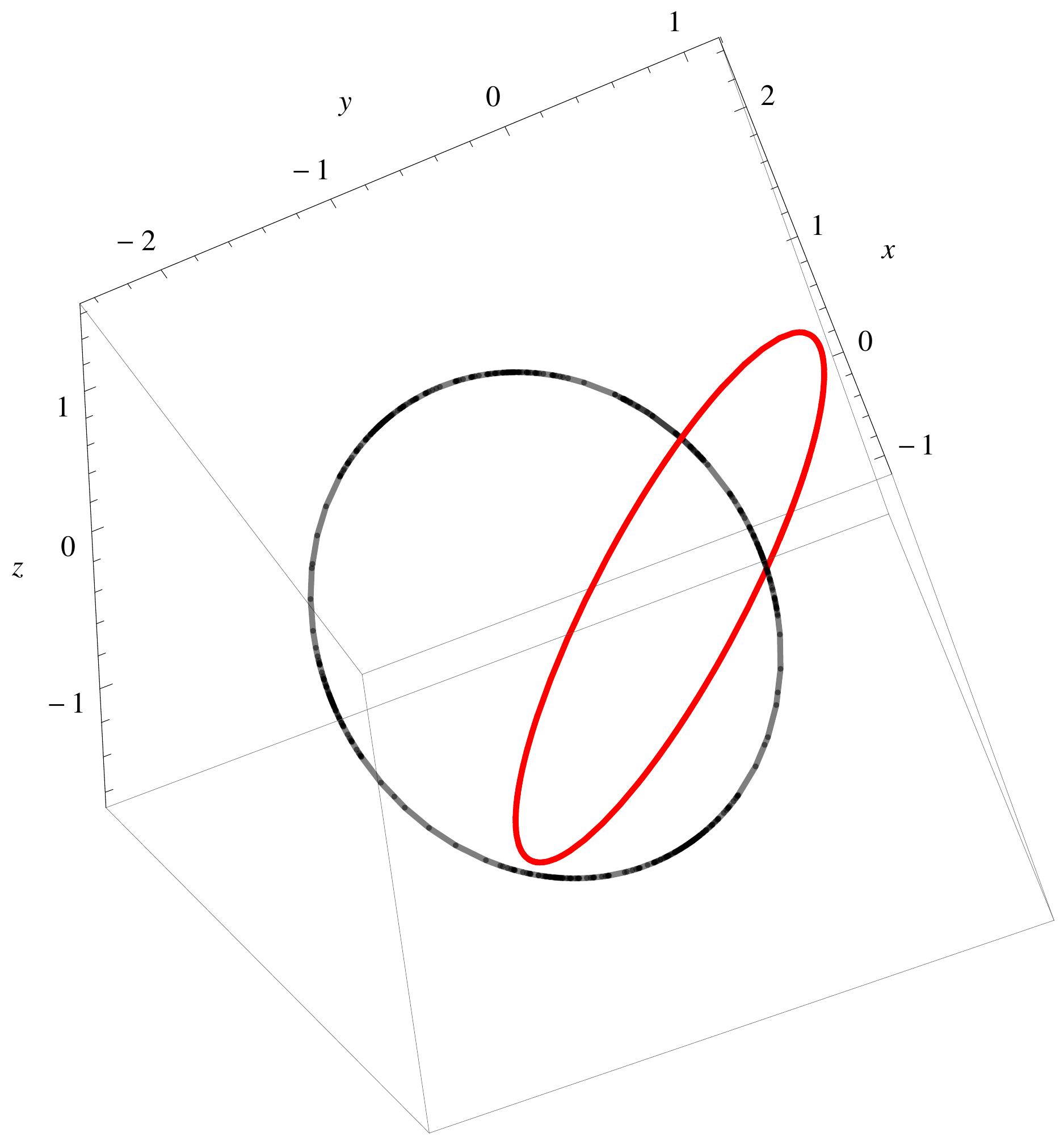}
        \caption{}
        \label{fig:HopfionEBFL}
    \end{subfigure}
    \hfill
    \begin{subfigure}[b]{0.4 \textwidth}
        \centering
        \includegraphics[width=\textwidth]{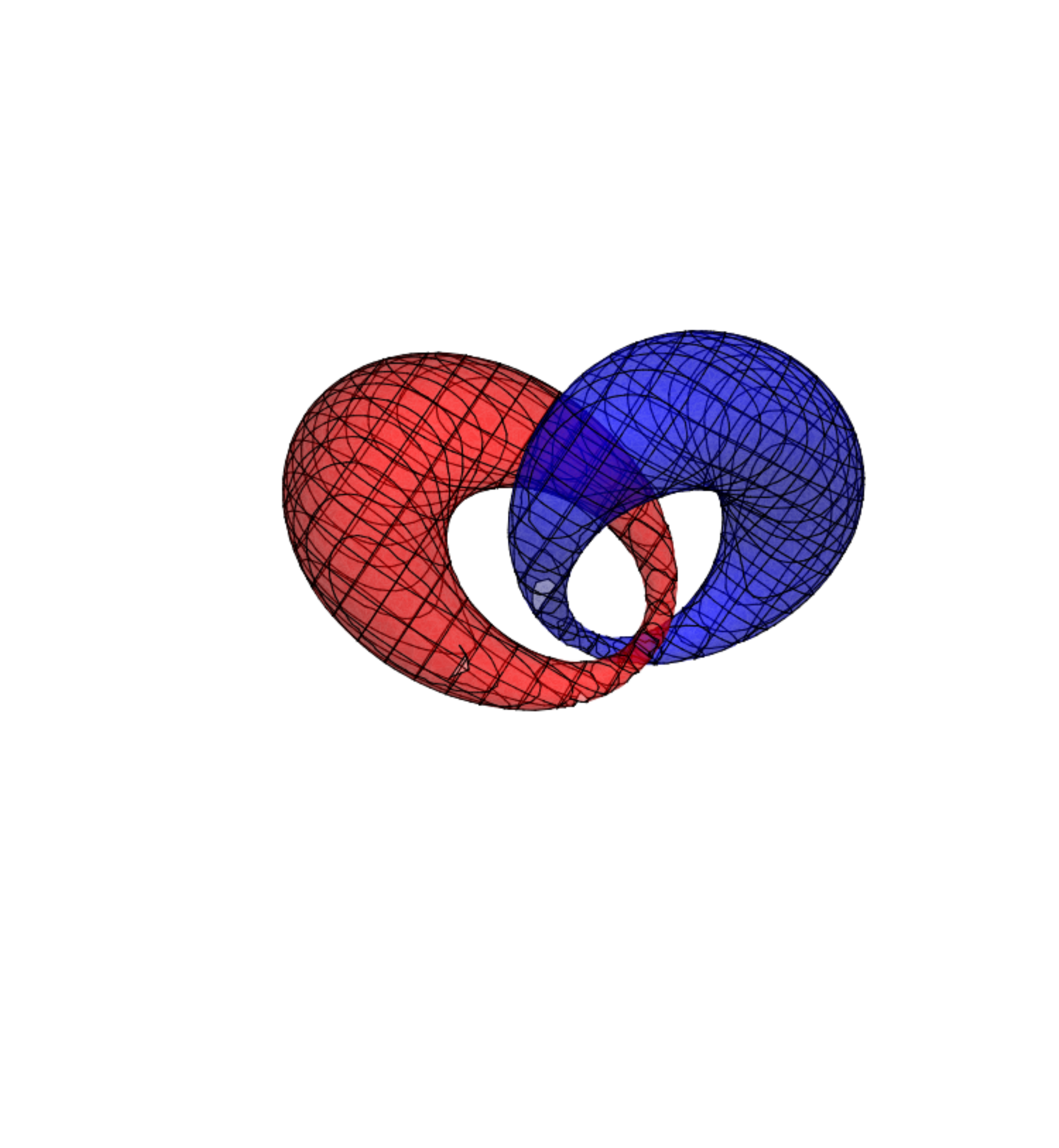}
        \caption{}
        \label{fig:HopfionEBFS}
    \end{subfigure}
    \begin{subfigure}[b]{0.4 \textwidth}
        \centering
        \includegraphics[width=\textwidth]{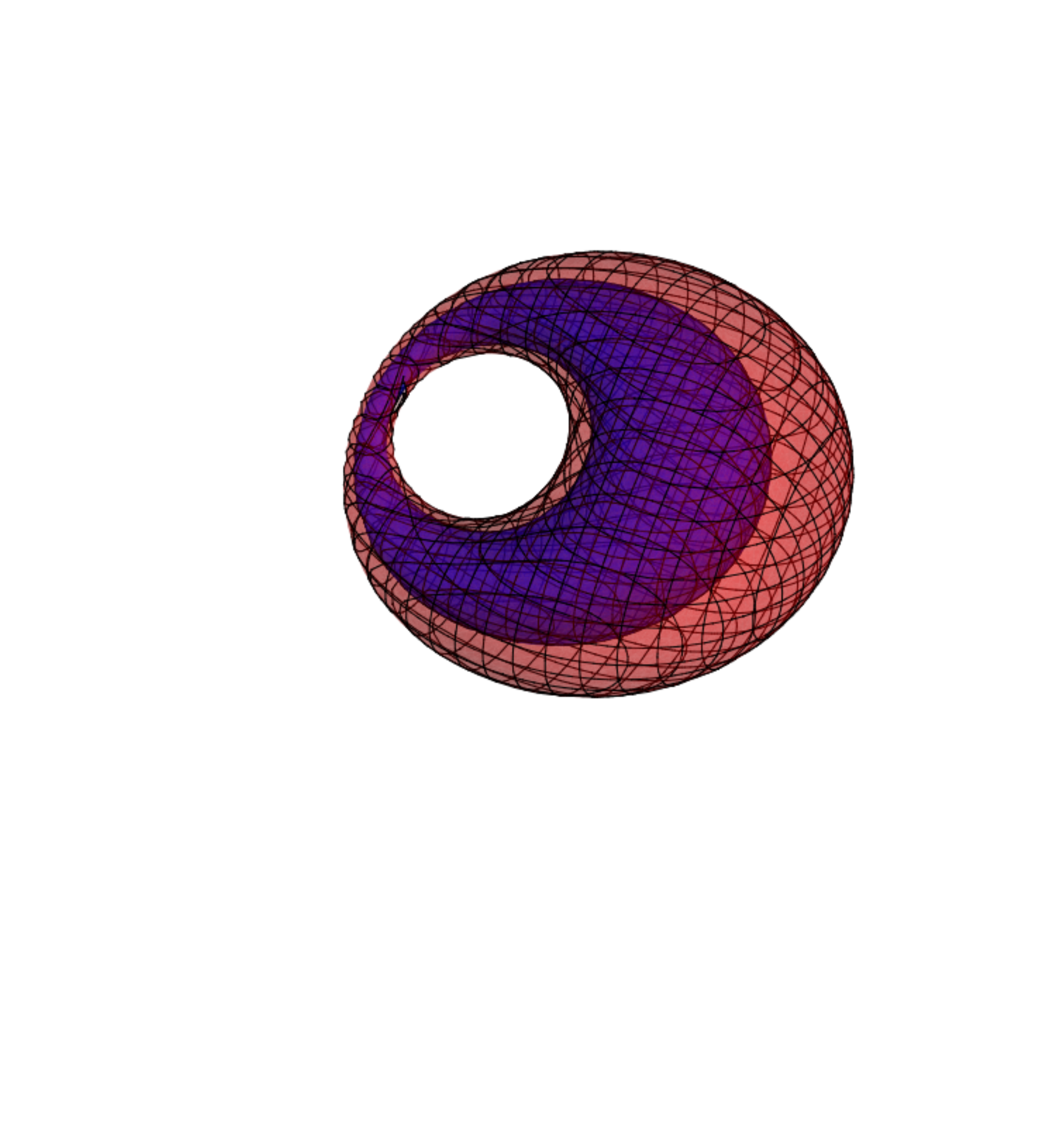}
        \caption{}
        \label{fig:HopfionEEFS}
    \end{subfigure}
    \caption{ Structure of Hopfion Solution.
    Fig(\ref{fig:HopfionEBFLdis}): Plot showing periodicity of field lines - electric (Black), magnetic (Red)
    (which are overlapping in the figure).  Fig.(\ref{fig:HopfionEBFL}):
   Shows electric field line (in Black) and magnetic field line (in Red) at $t=0$.
   Fig.(\ref{fig:HopfionEBFS}):Plot of surfaces $\phi_E=.45$ (Red) and $\phi_B=.45$ (Blue) at $t=0$.
   Fig.(\ref{fig:HopfionEEFS}):Plot of surfaces $\phi_B=.4$ (Red) and $\phi_B=.45$ (Blue) at $t=0$}
    \label{fig:Hopfion}
\end{figure}

The Hopfion solution has finite energy $E$, momentum $\mb{P}=(0,0,-E/2)$ and angular momentum $\mb{L}=(0,0,E/2)$. 
It also has finite helicities $h_{ee}=h_{mm}=E/2$. The Hopfion solution is topologically non-trivial in the sense that electric and magnetic field lines are linked with each other (Fig.(\ref{fig:HopfionEBFL})). 

The Hopfion solution was used in ~\cite{PhysRevLett.111.150404}  to generate a large family of topologically non-trivial solutions with different kinds of knots and links. 
The new solutions were found using the following observation: each pair $(\alpha,\beta)$ in Bateman's construction generates 
a whole family of solutions by simply defining new functions $(f,g)$ that depend holomorphically on $\alpha$ and $\beta$. 
The new field strength is changed to
\bea
\mb{F} &=& \nabla f(\alpha,\beta) \times \nabla g(\alpha,\beta)= h(\alpha,\beta) \nabla \alpha \times \nabla \beta,\\
\imath \left( \partial_t f \nabla g-\partial_t g \nabla f\right)&=& h(\alpha,\beta) \imath \left( \partial_t \alpha \nabla \beta-\partial_t \beta \nabla \alpha\right).
\eea
where $h=\partial_{\alpha} f \partial_{\beta} g-\partial_{\beta} f \partial_{\alpha} g $. Clearly, if $(\alpha,\beta)$ satisfy
the equations of motion, so do $(f,g)$. The new solutions have $f=\alpha^p$ and $g= \beta^q$.  For co-prime integers $(p,q)$ one finds various knots.
The $(p,q)$-knot solutions have finite energy $E$, that depends on the amplitude of the solutions. The other charges are
\begin{itemize}
\item Momentum: $\mb{P}=(0,0,- \frac{p}{p+q}E)$.
\item Angular momentum: $\mb{L}=(0,0,\frac{q}{p+q}E)$.
\item Boosts: $\mb{B}_v=(0,0,0)$.
\item Dilatations: $D=0$.
\item TSCT: $K^0=E$.
\item SSCT: $\mb{K}=(0,0,-\frac{p}{p+q}E)$
\item Helicities: $h_{ee}=h_{mm}=\frac{1}{p+q}E$.
\item Cross helicities: $h_{em}=-h_{me}=0$.
\end{itemize}

We can obtain the field lines corresponding to a vector field $V^i(x^i)$ by solving for curves given by  $x^i(\tau)$, where,
\be
\frac{d x^i(\tau)}{d \tau}= V^i(x^i(\tau)).
\ee
This first order differential equation, along with the boundary condition $x^i(\tau=0)=x^i_0$ gives a field line
passing through the point $x^i_0$.
For the case of  electric or magnetic lines, we use normalized electric or magnetic field at a given time as $V^i$. To check
the periodicity of these lines of force we can plot $\ln(|x^i(\tau)-x^i(0)|)$ as a function of $\tau$. For the Hopfion such a plot
is given in Fig.(\ref{fig:HopfionEBFLdis}). We also solve numerically field lines for the Hopfion and plot them in Fig.(\ref{fig:HopfionEBFL}).

Alternatively, the structure of the solutions can be studied using surfaces that contain electric and magnetic field lines  and that we define below.
For the $(p,q)$ solutions with $|\alpha|^2+|\beta|^2=1$, we can find a gradient field orthogonal to the electric and magnetic fields given by~\cite{PhysRevLett.111.150404},
\be\label{defphiEB}
\phi=\phi_B+ \imath \phi_E = \alpha^p \beta^q ~~;~~ \mb{B} \cdot \nabla \phi_B = \mb{E} \cdot \nabla \phi_E=0.
\ee
So we can visualize the geometrical properties of these solutions by plotting the isosurfaces of
$\phi_E$ and $\phi_B$, which contain the electric and magnetic field lines. Figs.~(\ref{fig:HopfionEBFS}) and (\ref{fig:HopfionEEFS}) shows such isosurfaces for the Hopfion.
Properties of these isosurfaces were studied in detail in \cite{PhysRevLett.111.150404}.

\section{Generating new solutions}

There are multiple possibilities to generate new solutions. One way that we have already explained is to 
substitute $\alpha$ and $\beta$ in Bateman's construction by holomorphic functions of the same solutions, 
with arbitrary complex coefficients. Another way is to use the symmetries of the equations to produce new solutions. 
Clearly if we do a Poincar\'e transformation we will trivially obtain a new solution. 
Translations will not affect to the charges of the solutions, while rotations will rotate momentum and angular momentum and 
boosts will also change the energy and conformal charges. However, we do not see them as truly new solutions, 
since we can recover the original form of the solution by simply changing the frame of reference.

A more interesting possibility is to consider other kind of deformations, such as scaling and squeezing, 
or conformal transformations~\cite{hep-th/9701064}. This is used sometimes in a different context in order to check 
the stability of soliton solutions (see e.g. \cite{Manton:2008ca,Domokos:2013xqa}). 
In general not all transformations will lead to a new solution, in order for this to happen it is necessary that the 
equations of motion are on-shell invariant. In the case of Bateman's construction there is a new interesting generalization. 
Since the equations are complex, the deformations can also be made complex\footnote{It was pointed out to us
(after appearance of our preprint in arXiv) that, previously complex transformation of Maxwell solutions was studied in
~\cite{trautman1962, Newman1973, bialynicki2004electromagnetic,bialynicki2013role}. See also \cite{Fouchtchitch87,DalhuisenThesis}.}. 
In practice this could be interpreted as mixing usual deformations with electromagnetic duality, 
that as we have seen corresponds to introducing imaginary factors in the solutions.

We will first use infinitesimal transformations to learn which ones can be used to generate new solutions. 
Then, we will do finite transformations of the solutions and study their properties. In all cases we have checked 
explicitly that the equations of motion are satisfied by the new deformed solutions. Although these complex
transformations on the solutions generate ``new'' solutions to the equations of motion, they might be sometimes
related to each other by a real transformation, and thus be in the same class.

For convenience we will use components rather than vectorial notation. Greek indices $\mu,\nu=0,1,2,3$ refer to spacetime directions and Latin indices $i,j=1,2,3$ refer to spatial directions. For a spatial vector the curl and the divergence are
\begin{equation}
(\nabla\times \mb{A})^i=\epsilon^{ijk}\partial_j A_k,\ \ (\nabla\cdot \mb{A})=\partial_i A^i.
\end{equation}
For four-vectors we will raise and lower indices with the mostly plus metric $\eta_{\mu\nu}=\text{diag}\,(-1,1,1,1)$ and define the epsilon tensor with spacetime indices as $\epsilon^{0ijk}=-\epsilon^{ijk}$. 

\subsection{Infinitesimal transformations of the solutions}

Under an infinitesimal  (complex) coordinate transformation $x^\mu\to x^\mu+\xi^\mu$, the functions $\alpha$ and $\beta$ change as
\begin{equation}
\delta\alpha=\xi^\lambda\partial_\lambda\alpha, \ \ \delta\beta = \xi^\lambda\partial_\lambda\beta.
\end{equation} 
The equation of motion is
\begin{equation}
i(\partial_0\alpha\partial^i\beta-\partial^i\alpha\partial_0\beta)-\epsilon^{ijk}\partial_j\alpha\partial_k\beta=0.
\end{equation}
Under the infinitesimal transformation there is a part $\sim \xi^\lambda\partial_\lambda(\text{equation})$ which vanishes trivially on-shell, and then there are the contributions
\begin{equation}
0=i\partial_0\xi^\lambda(\partial_\lambda\alpha\partial^i\beta-\partial^i\alpha\partial_\lambda\beta)+i\partial^i\xi^\lambda(\partial_0\alpha\partial_\lambda\beta-\partial_\lambda\alpha\partial_0\beta)-\partial_j\xi^\lambda\epsilon^{ijk}(\partial_\lambda\alpha\partial_k\beta-\partial_k\alpha\partial_\lambda \beta).
\end{equation}
We can write each term as follows using the equations of motion
\begin{eqnarray}
i\partial_0\xi^\lambda(\partial_\lambda\alpha\partial^i\beta-\partial^i\alpha\partial_\lambda\beta) &=& \left(\partial_0\xi^0\delta^{i}_l+i\partial_0 \xi^n \epsilon_{ln}^{\ \ i}\right)\epsilon^{ljk}\partial_j\alpha\partial_k\beta,\nonumber \\
i\partial^i\xi^\lambda(\partial_0\alpha\partial_\lambda\beta-\partial_\lambda\alpha\partial_0\beta)& =& \partial^i\xi^l \epsilon_l^{\ jk}\partial_j\alpha\partial_k\beta,\nonumber \\
\partial_j\xi^\lambda\epsilon^{ijk}(\partial_\lambda\alpha\partial_k\beta-\partial_k\alpha\partial_\lambda \beta) &=& \left(i\partial_n\xi^0\epsilon_{l}^{\ n i}-\partial_l\xi^i+\partial_n\xi^n\delta_l^i\right)\epsilon^{ljk}\partial_j\alpha\partial_k\beta.
\end{eqnarray}
All together, we find the condition
\begin{equation}
0=\left[ (\partial_0\xi^0-\partial_n\xi^n)\delta^{il}+i(\partial_0\xi^n-\partial_n\xi^0)\epsilon^{lni}+\partial^i\xi^l+\partial^l \xi^i\right]\epsilon_l^{\ jk}\partial_j\alpha\partial_k\beta.
\end{equation}
It is immediate to see that this is satisfied for translations in all coordinates $\xi^\mu=c^\mu$, Lorentz transformations $\xi^\mu=\Lambda^\mu_{ \ \nu}x^\nu$, with $\Lambda_{\mu\nu}=-\Lambda_{\nu\mu}$ and dilatations $\xi^\mu= \lambda x^\mu$. 

Let us write the derivative in the general form
\begin{equation}
\partial_i \xi_j= \frac{1}{3}\delta_{ij}\partial_n\xi^n+\frac{1}{2}\sigma_{ij}+\frac{1}{2}\omega_{ij},
\end{equation}
where $\omega_{ij}$ is the antisymmetric part and $\sigma_{ij}$ is symmetric and traceless. Then,
\begin{equation}
0=\left[ \left(\partial_0\xi^0-\frac{1}{3}\partial_n\xi^n\right)\delta^{il}+i(\partial_0\xi^n-\partial_n\xi^0)\epsilon^{lni}+\sigma^{il}\right]\epsilon_l^{\ jk}\partial_j\alpha\partial_k\beta.
\end{equation} 
We see that the following conditions should be satisfied in general:
\begin{equation}
\sigma_{ij}=0,\ \ \partial_0\xi^0-\frac{1}{3}\partial_n\xi^n=0,\ \ \partial_0\xi^n-\partial_n\xi^0=0.
\end{equation}
Note that these conditions are also satisfied by special conformal transformations
\begin{equation}
\xi^\mu= a^\mu x^2-2a_\lambda x^\lambda x^\mu.
\end{equation}
So new solutions can be generated by using conformal transformations with complex parameters. Shear deformations on the other hand do not produce new solutions.

\subsection{(p,q)-knots from constant electric and magnetic field}

As a very simple application of  complex deformations let us show how the Hopfion solution or (1,1)-knot 
can be obtained through a 
conformal transformation of a solution with constant electric and magnetic fields.

Let us  choose the solution with constant field strength as given in \eqref{abconstantF},
\be
\alpha= 2 \imath (t+z)-1 ~~;~~ \beta = 2 ( x- \imath y) 
\ee
The corresponding electric and magnetic field are constant
and  have diverging total energy and helicity. Now let us consider a special conformal transformation (SCT),
\bea\label{conftrans}
x^{\mu} \to \frac{x^{\mu} - b^{\mu} x_\sigma x^\sigma}{1- 2 b_\sigma x^\sigma+ b_\sigma b^\sigma x_\rho x^\rho}.
\eea
With the choice of the parameter, $b^{\mu} = \imath (1,0,0,0)$ the new $(\alpha,\beta)$ obtained by this 
transformation is the same as the Hopfion solution given by \eqref{hopfalpha1} and \eqref{hopfbeta1}
~\footnote{It was pointed out to us (after appearance of our preprint in arXiv) that, in ~\cite{bialynicki2004electromagnetic},
it was shown that similar linked solution of Maxwell equations can be obtained from constant solutions
by a coordinate transformation given by  conformal reflection
accompanied by the shift in time direction by an imaginary constant. See also \cite{DalhuisenThesis}.}. 
If we consider more general choice of the parameter $b^{\mu}$, we can generate linked solution with various
values of the conserved charges.

Other knotted solutions can be generated by considering integer powers of $(\alpha, \beta)$ given in \eqref{abconstantF},
\ie $(\alpha, \beta) \to (\alpha^p, \beta^q)$, and then
considering a SCT with parameter $b^{\mu} = \imath (1,0,0,0)$. Fig.(\ref{fig:(2,3)knot}) shows
a isosurface of $\phi_E$ (as defined in \eqref{defphiEB}) for $(p=2,q=3)$-knot. 
\begin{figure}
\center
 \includegraphics[scale=.5]{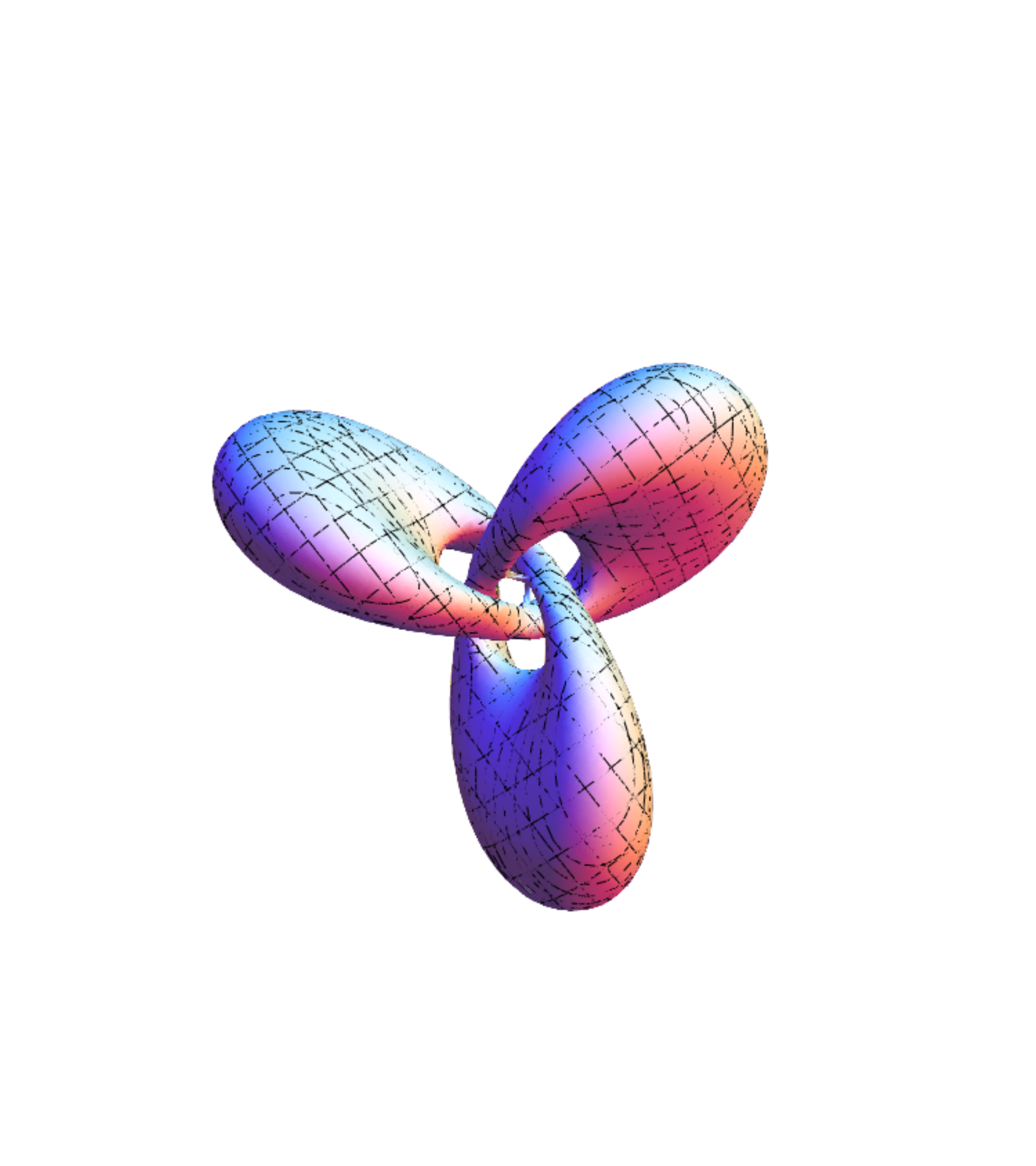}
 \caption{Plot of $\phi_E=0.05$ surface for $(2,3)$-knot at $t=0$}\label{fig:(2,3)knot}
\end{figure}
The final $(\alpha,\beta)$ obtained here is same as that given in
 section (\ref{sec:knownsolution}) for $(p,q)$ knots. So generating new solution by
 action of SCT and considering holomorphic functions of solution commutes
 with each other. The solution before considering the SCT, given by $(\alpha^p, \beta^q)$,
 has non-constant electric and magnetic fields with divergent total energy and helicity. 
\subsection{Knotted solutions from plane waves}

We can use complex transformations also to generate new solutions from the plane wave \eqref{PWalphabeta}.
Let us write here again the solution,
\be
\alpha = e^{\imath (z-t)} ~~;~~ \beta = x + \imath y. 
\ee
We can easily notice that not all complex transformations
will generate new solutions. An imaginary translation in $x$ or $y$ is equivalent to a real space translation along $y$ or $x$
respectively.
An imaginary translation in $z$ or $t$ is equivalent to the same solution (\ie~field strength) up to an overall scale.

\begin{figure}
    \centering
    \begin{subfigure}[b]{0.3 \textwidth}
        \centering
        \includegraphics[width=\textwidth]{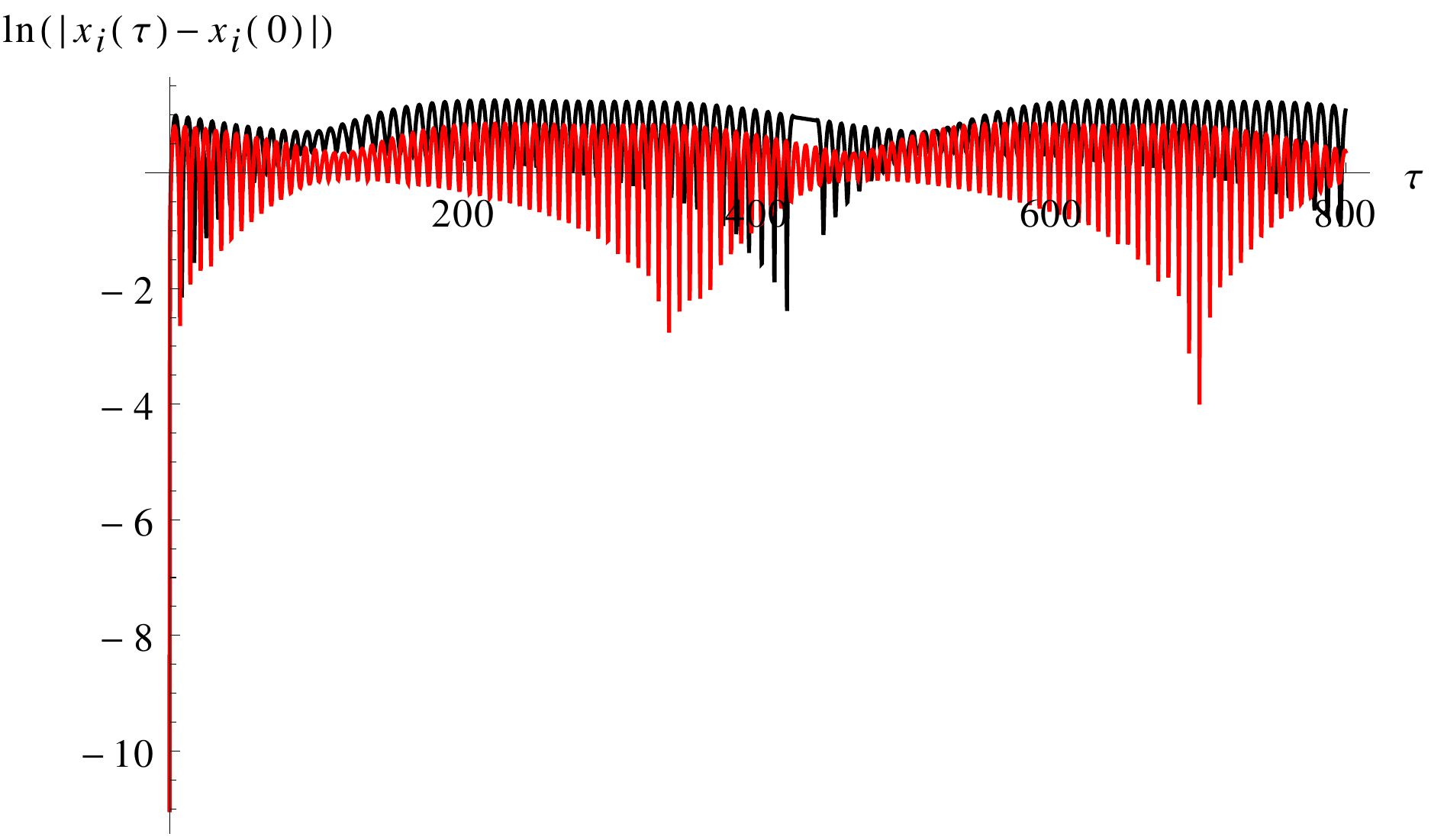}
        \caption{}
        \label{fig:PWSCTEBFLdis}
    \end{subfigure}
    \hfill
    \begin{subfigure}[b]{0.3 \textwidth}
        \centering
        \includegraphics[width=\textwidth]{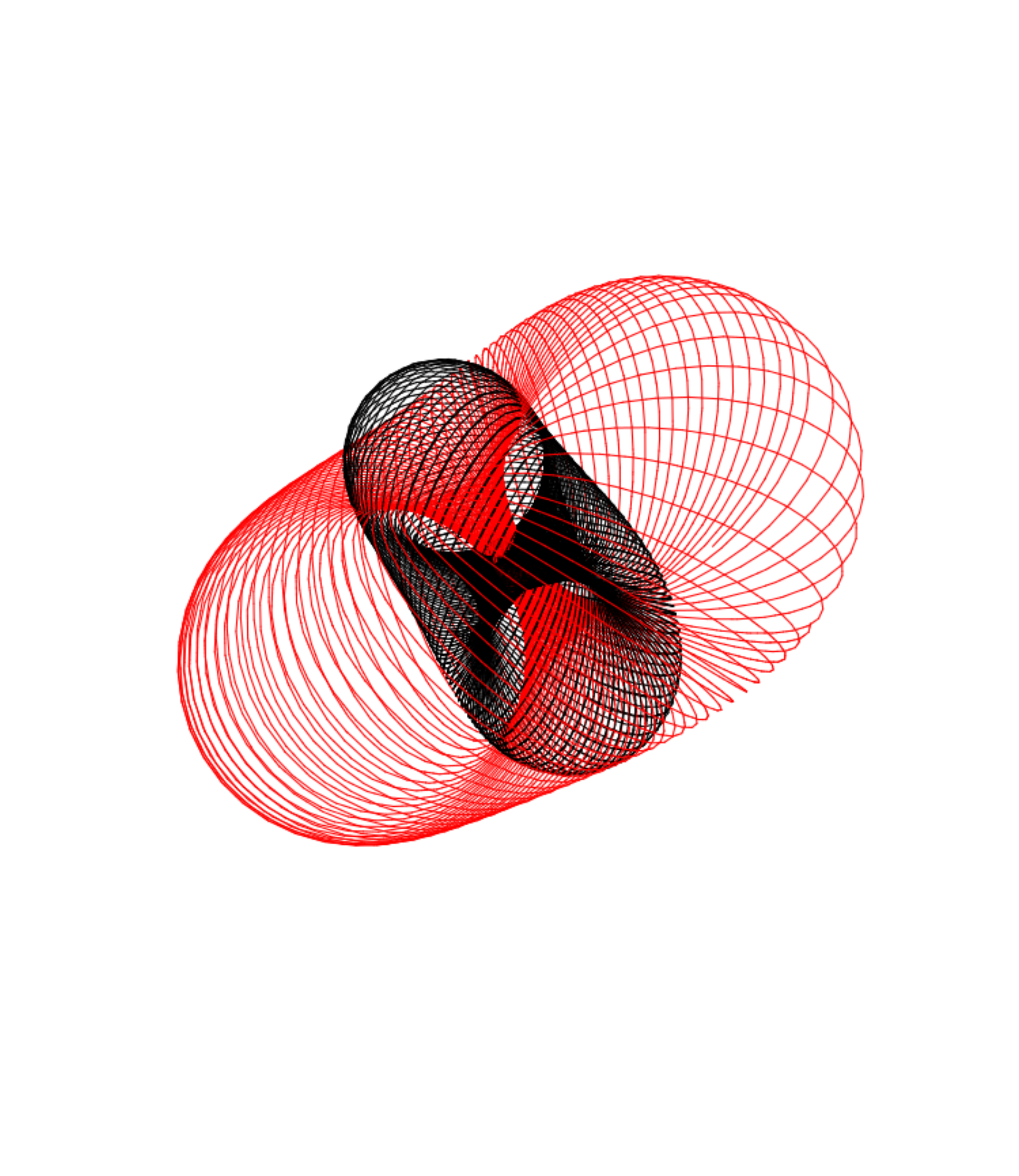}
        \caption{}
        \label{fig:PWSCTEBFL}
    \end{subfigure}
    \hfill
    \begin{subfigure}[b]{0.3 \textwidth}
        \centering
        \includegraphics[width=\textwidth]{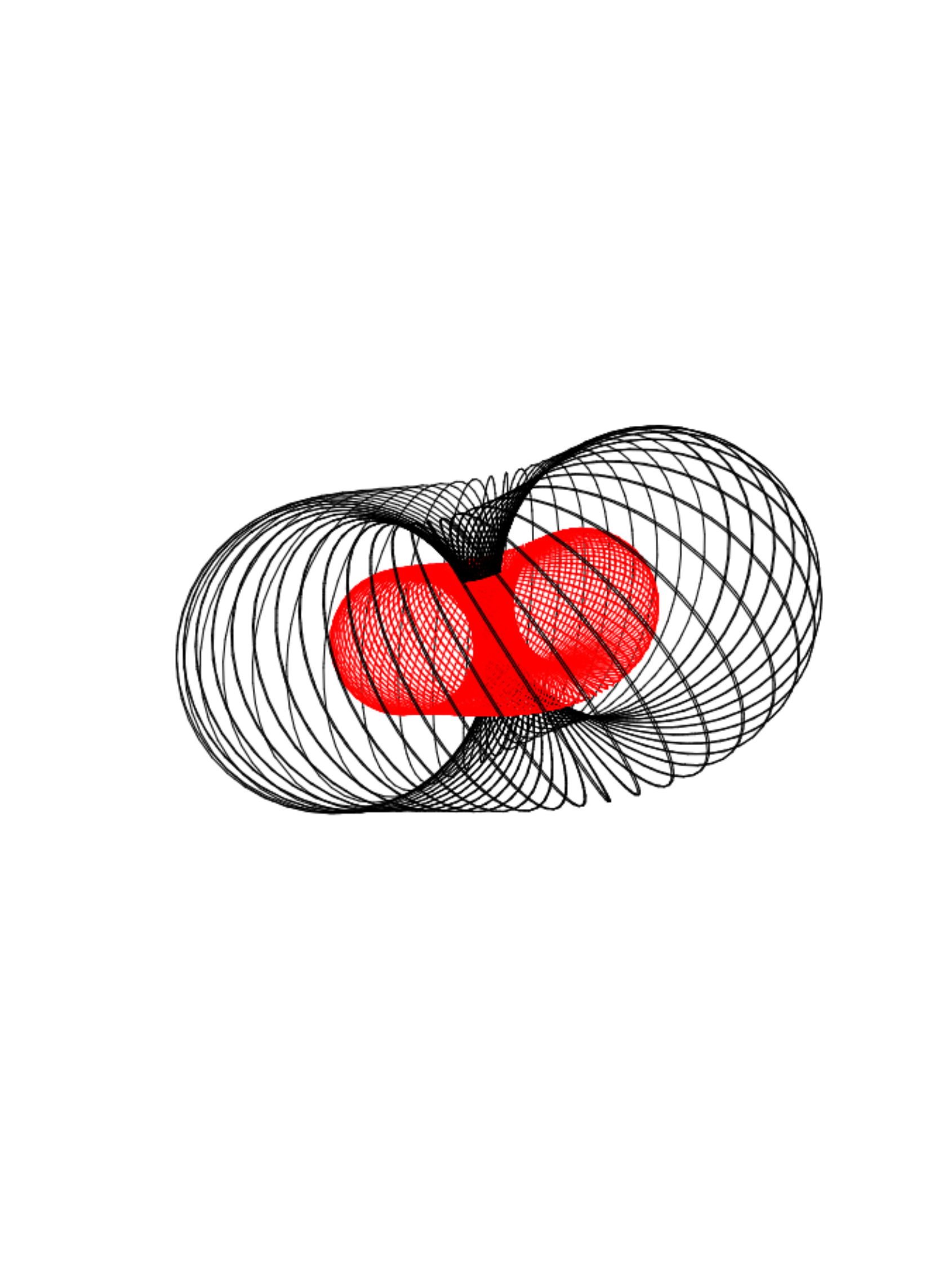}
        \caption{}
        \label{fig:PWSCTEEFL}
    \end{subfigure}
    
    \caption{Knotted Solutions from Plane Waves. 
    Fig(\ref{fig:PWSCTEBFLdis}): Plot showing periodicity of field lines - electric (Black), magnetic (Red). 
    Fig.(\ref{fig:PWSCTEBFL}):
    Shows an electric field line (in Black) and a magnetic field line (in Red) at $t=0$.
   Fig.(\ref{fig:PWSCTEEFL}): Shows two  electric field lines  at $t=0$}
    \label{fig:PWSCT}
\end{figure}

 The only  transformations that are left to produce new solutions are special conformal transformations given by \eqref{conftrans}.
Let us choose $b^{\mu}$ as $(\imath,0,0,0)$. The new solution for $\alpha,\beta$ is given by,
\bea
\alpha &=& e^{(-1+\frac{\imath (t+z-\imath)}{1-t^2+x^2+y^2+z^2+ 2 \imath t})},\label{PTKalpha}\\
\beta &=& \frac{x+\imath y}{1-t^2+x^2+y^2+z^2+ 2 \imath t}.\label{PTKbeta}
\eea
Note that the solution is smooth. This is not true in general for real conformal transformations.
In contrast to the plane wave solution, this solution has finite energy $E$ and the following finite conserved charges fixing $E=1$ 
\begin{itemize}
\item Momentum $\mb{P}=(0,0,0.613)$.
\item Angular momentum $\mb{L}=(0,0,-0.387)$.
\item Boosts: $\mb{B}_v=(0,0,0)$.
\item Dilatations: $D=0$.
\item TSCT: $K^0=0.613$.
\item SSCT:  $\mb{K}=(0,0,0.225)$.
\item Helicities $h_{ee}=h_{mm}=0.387$.
\item Cross helicities $h_{em}=-h_{me}=0$.
\end{itemize}
A plot of the field lines (Fig.\ref{fig:PWSCT}) shows that the electric and magnetic field lines lie on a torus and are linked to each other, so the solution is topologically non-trivial.

\subsection{Complex deformations of Hopfions}

For the Hopfions we find that all complex transformations lead to new solutions. We list the form of the new solutions and the conserved charges below:

\begin{itemize}
\item {\bf Time translation:}\\
The solutions generated by the imaginary time translation $t\to t+\imath c$, with $c$ a constant real parameter  are,
\bea
\alpha &=& N_1 \frac{A-1+ \imath z}{A+  \imath (t+ \imath c)}, \label{newalpha1} \\
\beta &=& N_1 \frac{(x-\imath y) }{A+ \imath (t+ \imath c)}.\label{newbeta1}
\eea
where $A=\half(x^2+y^2+z^2-(t+ \imath c)^2+1)$. We introduce a factor $N_1$ as the overall normalization, which is related to the energy $E$ as
$N_1^4=\frac{E |1-c|^5}{2 \pi^2}$. 

The conserved charges for this solution are
\begin{itemize}
\item Momentum: $\mb{P}=(0,0,- \frac{E}{2})$.
\item Angular momentum: $\mb{L}=(0,0,\frac{1-c}{2}E)$.
\item Boosts: $\mb{B}_v=(0,0,0)$.
\item Dilatations: $D=0$.
\item TSCT: $K^0=(1-c)^2 E$.
\item SSCT: $\mb{K}=(0,0,-\frac{1}{2}(1-c)^2E)$.
\item Helicities: $h_{ee}=h_{mm}=\frac{1-c}{2}E$.
\item Cross helicities : $h_{em}=-h_{me}=0$
\end{itemize}
The helicity in units of energy is a continuous function of the parameter $c$, but the
structure of electric and magnetic lines are same as that of Hopfion for all values of $c\neq 1$.

\item {\bf Space translation:}\\
Deformed solutions have generically divergent energy except for a finite range of the translation parameter. Let us consider the case where only the $z$ direction is shifted $z \to z + \imath c$, 
with $c$ real constant number.
\bea
\alpha &=& N_1 \frac{A-1+ \imath (z+ \imath c)}{A+  \imath t} \label{newalpha2}, \\
\beta &=& N_1 \frac{(x-\imath y) }{A+ \imath t}\label{newbeta2}.
\eea
where $A=\half(x^2+y^2+(z+ \imath c)^2-t^2+1)$.
In order to have finite energy solutions we are forced to require $|c|<1$. 
It is difficult to find an analytic expression for charges as function of the parameter. Instead, Figure~\ref{fig:HelSpaceComplex} shows a plot of
helicity vs the parameter $c$ in units such that total energy is $E=1$.
\begin{figure}[thb!]
\begin{center}
 \includegraphics[width=8cm]{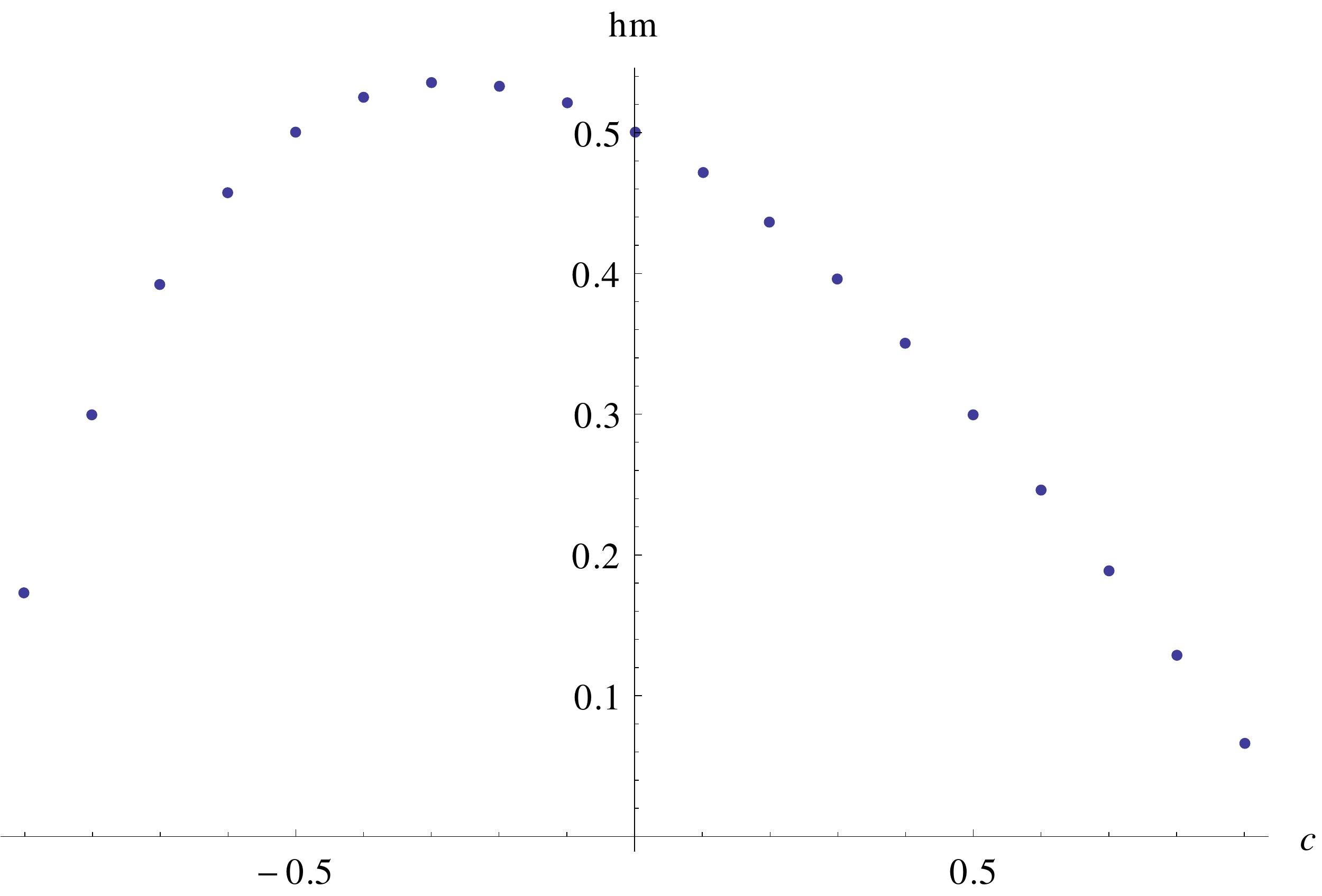}
 \caption{Helicity  (in units  total energy $=1$)  vs $c$ for a solution generated by a complex spatial translation.}
 \label{fig:HelSpaceComplex}
\end{center}
\end{figure}
We have computed numerically the charges for $c=0.5$ and $E=1$:
\begin{itemize}
\item Momentum: $\mb{P}=(0,0,-0.5)$.
\item Angular momentum: $\mb{L}=(0,0,0.3)$.
\item Boosts: $\mb{B}_v=(0,0,0)$.
\item Dilatations: $D=0$.
\item TSCT: $K^0=0.45$.
\item SSCT: $\mb{K}=(0,0,-0.25)$.
\item Helicity: $h_{ee}=h_{mm}=0.3$.
\item Cross helicities : $h_{em}=-h_{me}=0$
\end{itemize}

\item {\bf Rotations:}\\
We can also generate new solutions using rotations along the spatial coordinates with complex angles. Rotations in the $x,y$ plane
by a complex angle amount to a rescaling of the solution, thus they do not generate really new solutions. Non-trivial transformations involve the $z$ direction. Let us consider the following  transformation in the $x,z$ plane, 
$x \to x \cos(\imath \theta)-z \sin( \imath \theta), ~ z \to z \cos(\imath \theta)+ x \sin(\imath \theta)$. 
The new solution produced by this transformation is,
\bea
\alpha &=& N_1 \frac{A-1+ \imath (z \cos(\imath \theta)+ x \sin(\imath \theta))}{A+  \imath t},\label{newalpha3}\\
\beta &=&  N_1 \frac{(x \cos(\imath \theta)-z \sin( \imath \theta)-\imath y) }{A+ \imath t}.\label{newbeta3}
\eea
where $A=\half(x^2+y^2+z^2-t^2+1)$.
The normalization$N_1$ is related to the energy $E$ as $N_1^4=\frac{E}{2 \pi^2 \cosh^2(\theta)}$.
The charges for these solutions are
\begin{itemize}
\item Momentum: $\mb{P}=(0,\half \tanh(\theta),- \half \text{sech}(\theta))E$.
\item Angular momentum: $\mb{L}=(0,-\half \tanh(\theta),\half \text{sech}(\theta))E$.
\item Boosts: $\mb{B}_v=(0,0,0)$.
\item Dilatations: $D=0$.
\item TSCT: $K^0=E$
\item SSCT: $\mb{K}=(0,\half \tanh(\theta),- \half \text{sech}(\theta))E$.
\item Helicities: $h_{ee}=h_{mm}=E/2$.
\item Cross helicities : $h_{em}=-h_{me}=0$
\end{itemize}
In this case the helicity in units of the energy does not change.

\item {\bf Boosts:}\\
An imaginary boost along $x^i$ direction is given by, 
\bea
t &\to& t \cosh (\imath \theta) - x^i \sinh (\imath \theta), \\
x^i &\to& x^i \cosh (\imath \theta) - t \sinh (\imath \theta),
\eea
where $\theta$ is real. The expression of $\alpha, \beta$ for Hopfions is rotationally symmetric in the $(x,y)$ plane, so let us consider imaginary
boosts along $x$ and $z$ here.  We also introduce an overall  normalization $N_1$ as before, 
related to the energy $E$ as $N_1^4=\frac{E}{2 \pi^2} |\cos \theta|^5$, independently of the direction
of the boost.

The charges for the solution generated by boost along $x$ direction (of energy $E$),
\begin{itemize}
\item Momentum: $\mb{P}=(0,\frac{E}{2} \sin \theta,-\frac{E}{2} \cos \theta)$.
\item Angular momentum: $\mb{L}=(0,0,\half E)$.
\item Boosts: $\mb{B}_v=(-\sin \theta E,0,0)$.
\item Dilatations: $D=0$.
\item TSCT: $K^0=E$.
\item SSCT: $\mb{K}=(0,-\half \sin \theta E,-\half \cos \theta E)$.
\item Helicities: $h_{ee}=h_{mm}=\frac{1}{2}\cos\theta E$.
\item Cross helicities: $h_{em}=-h_{me}=0$.
\end{itemize}

The charges for a solution generated by boost along $z$ (of energy $E$) are
\begin{itemize}
\item Momentum: $\mb{P}=(0,0,-\frac{E}{2})$.
\item Angular momentum: $\mb{L}=(0,0,\half \cos\theta E)$.
\item Boosts: $\mb{B}_v=(0,0,-\sin \theta E)$.
\item Dilatations: $D=\half \sin \theta E$.
\item TSCT: $K^0=E$.
\item SSCT: $\mb{K}=(0,0,-\half E)$.
\item Helicities: $h_{ee}=h_{mm}=\frac{1}{2}\cos\theta E$.
\item Cross helicities: $h_{em}=-h_{me}=0$.
\end{itemize}

\item {\bf Scaling:}\\
Complex scalings $x^{\mu} \to b x^{\mu}$ generate new solutions, but if $b$ is purely imaginary the energy of the new solutions
is always infinite and the helicity is zero. Instead is better to consider transformations with $b= 1+ \imath c$, where $c$
is real. Such solutions have finite and non-zero energy and helicity  for a finite range of the parameter $c$. For these solutions
it is difficult to get the analytic expression of the charges as function of $c$, but we can easily obtain numerical results. For $c=1$ and $E=1$ we find the following values:
\begin{itemize}
\item Momentum: $\mb{P}=(0,0,-0.5)$.
\item Angular momentum: $\mb{L}=(0,0,0.25)$.
\item Boosts: $\mb{B}_v=(0,0,0.25)$.
\item Dilatations: $D=-0.5$.
\item TSCT: $K^0=0.5$
\item SSCT: $\mb{K}=(0,0,-0.25)$.
\item Helicities: $h_{ee}=h_{mm}=0.25$.
\item Cross helicities: $h_{em}=-h_{me}=0$.
\end{itemize}

\item {\bf Conformal transformations:}\\
We used the conformal transformation \eqref{conftrans} to generate new solutions from the plane wave. The action of such transformation
on the Hopfion solution can generate new solutions with finite energy and helicity for some finite range of values of 
the parameter of the transformation. Let us consider $b^{\mu}$ as $(0,\imath \half,0,0)$. 
We can compute the charges numerically for $E=1$
\begin{itemize}
\item Momentum: $\mb{P}=(0.286,0,-0.429)$.
\item Angular momentum: $\mb{L}=(0.286,0,0.571)$.
\item Boosts: $\mb{B}_v=(0,0.286,0)$.
\item Dilatations: $D=0$.
\item TSCT: $K^0=1.714$
\item SSCT: $\mb{K}=(-1.143,0,-0.5714)$.
\item Helicities: $h_{ee}=h_{mm}=0.571$.
\item Cross helicities: $h_{em}=-h_{me}=0$.
\end{itemize}
Fig.(\ref{fig:HopfionSCTEBFL}) shows the electric and magnetic field lines for this solution. We can see that the lines 
are still linked once like Hopfions, but they are deformed. 
Fig.(\ref{fig:HopfionSCTDP}) gives a comparison of  the energy density (in a plane orthogonal to total momentum) at various time for the Hopfion and
the solution found from performing a SCT on the Hopfion with $b^{\mu}$ as $(0,\imath \half,0,0)$. This comparison clearly illustrates the different
nature  of the new solution and the Hopfion.

\begin{figure}
    \centering
    \begin{subfigure}[b]{0.4 \textwidth}
        \centering
        \includegraphics[width=\textwidth]{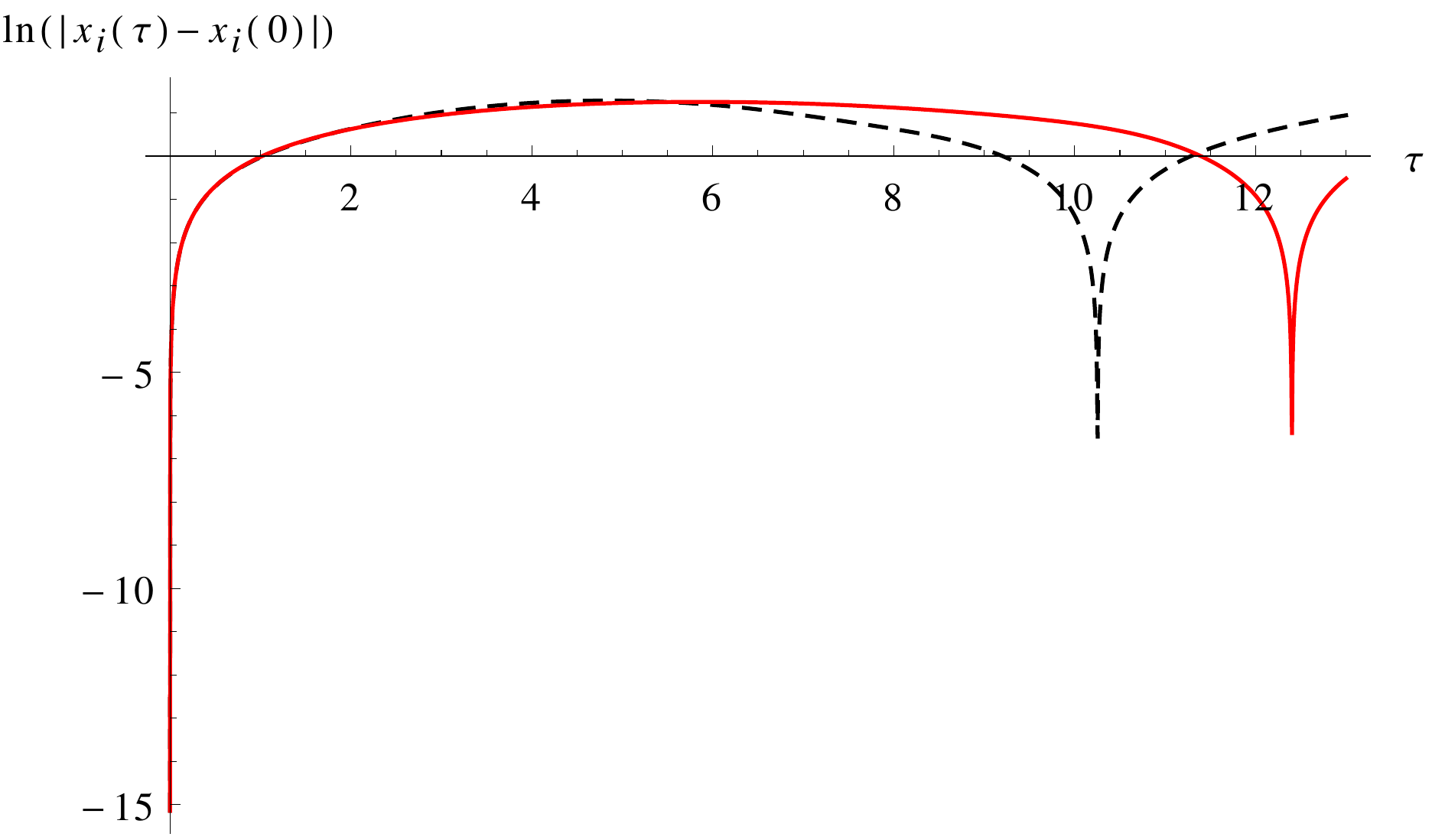}
        \caption{}
        \label{fig:HopfionSCTEBFLdis}
    \end{subfigure}
    \hfill
    \centering
    \begin{subfigure}[b]{0.4 \textwidth}
        \centering
        \includegraphics[width=\textwidth]{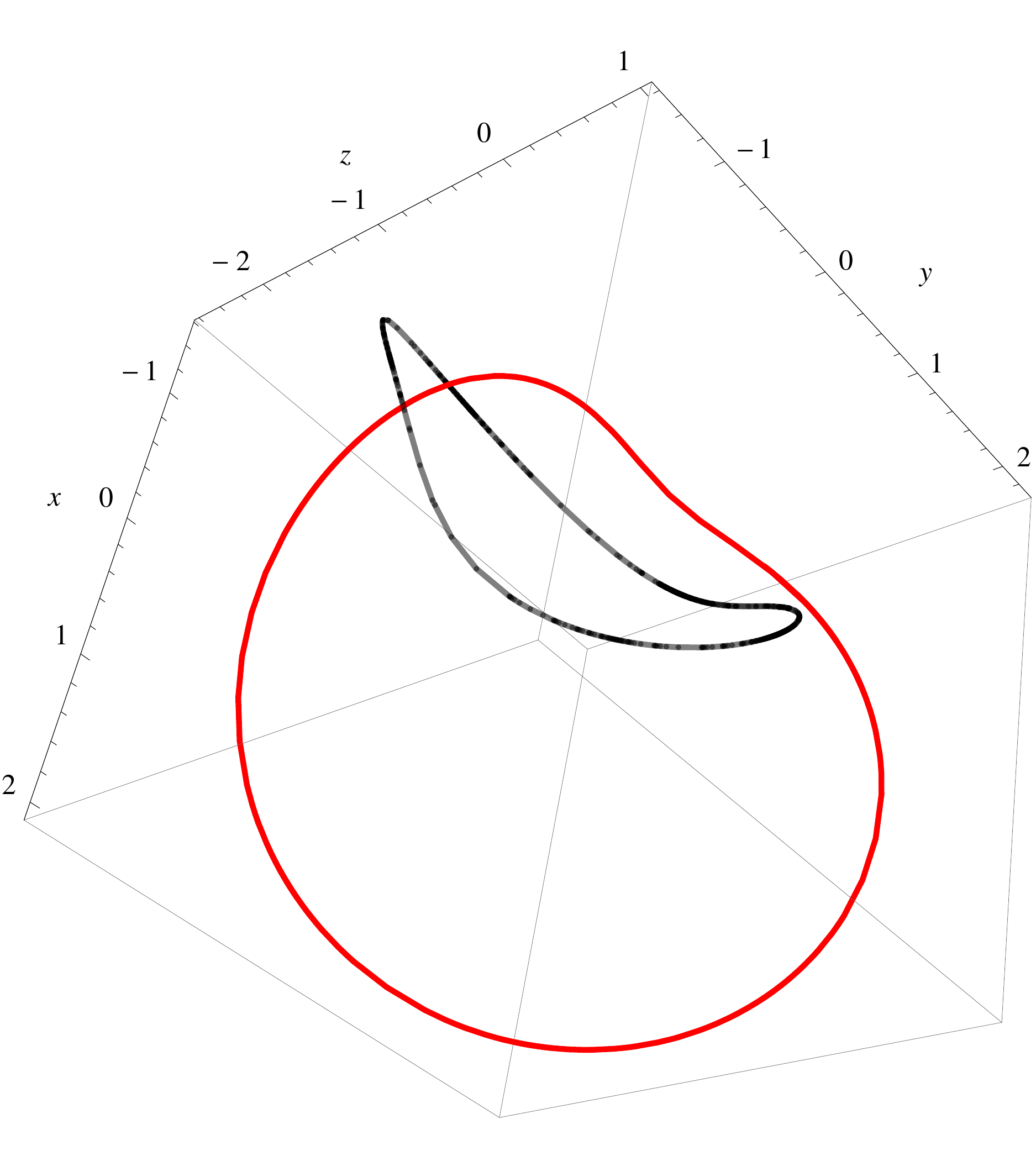}
        \caption{}
        \label{fig:HopfionSCTEBFL}
    \end{subfigure}
    \caption{Solution generated by SCT on Hopfion.
    Fig(\ref{fig:HopfionSCTEBFLdis}): Plot showing periodicity of field lines - electric (Black), magnetic (Red). 
    Fig.(\ref{fig:HopfionSCTEBFL}):
   Shows an electric field line (in Black) and a magnetic field line (in Red) at $t=0$.
   }
    \label{fig:HopfionSCT}
\end{figure}

\begin{figure}
\centering
    \begin{subfigure}[b]{0.2 \textwidth}
        \centering
        \includegraphics[width=\textwidth]{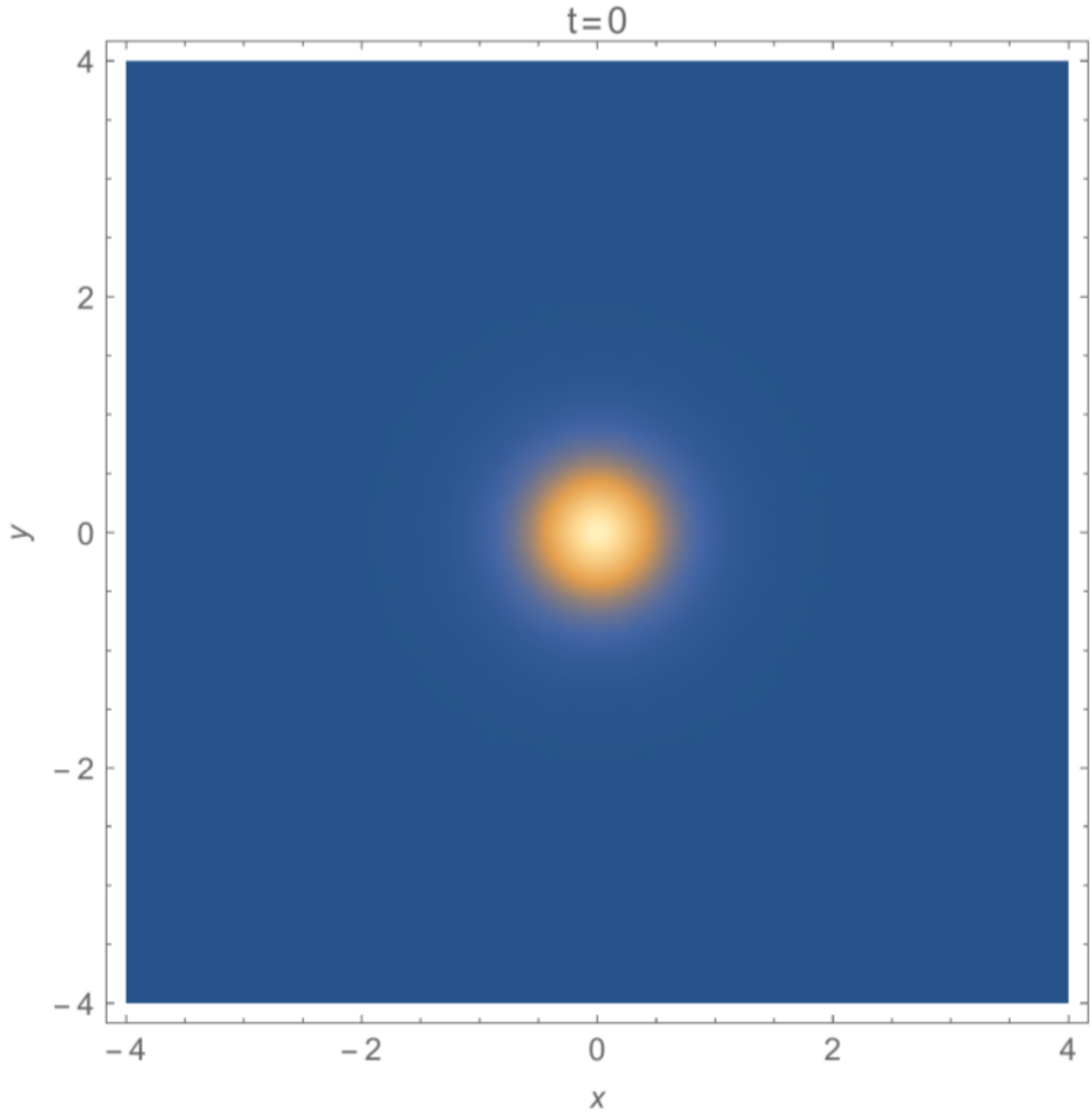}
        \caption{}
        \label{fig:HopfionDPt0}
    \end{subfigure}
    \hfill
    \centering
    \begin{subfigure}[b]{0.2 \textwidth}
        \centering
        \includegraphics[width=\textwidth]{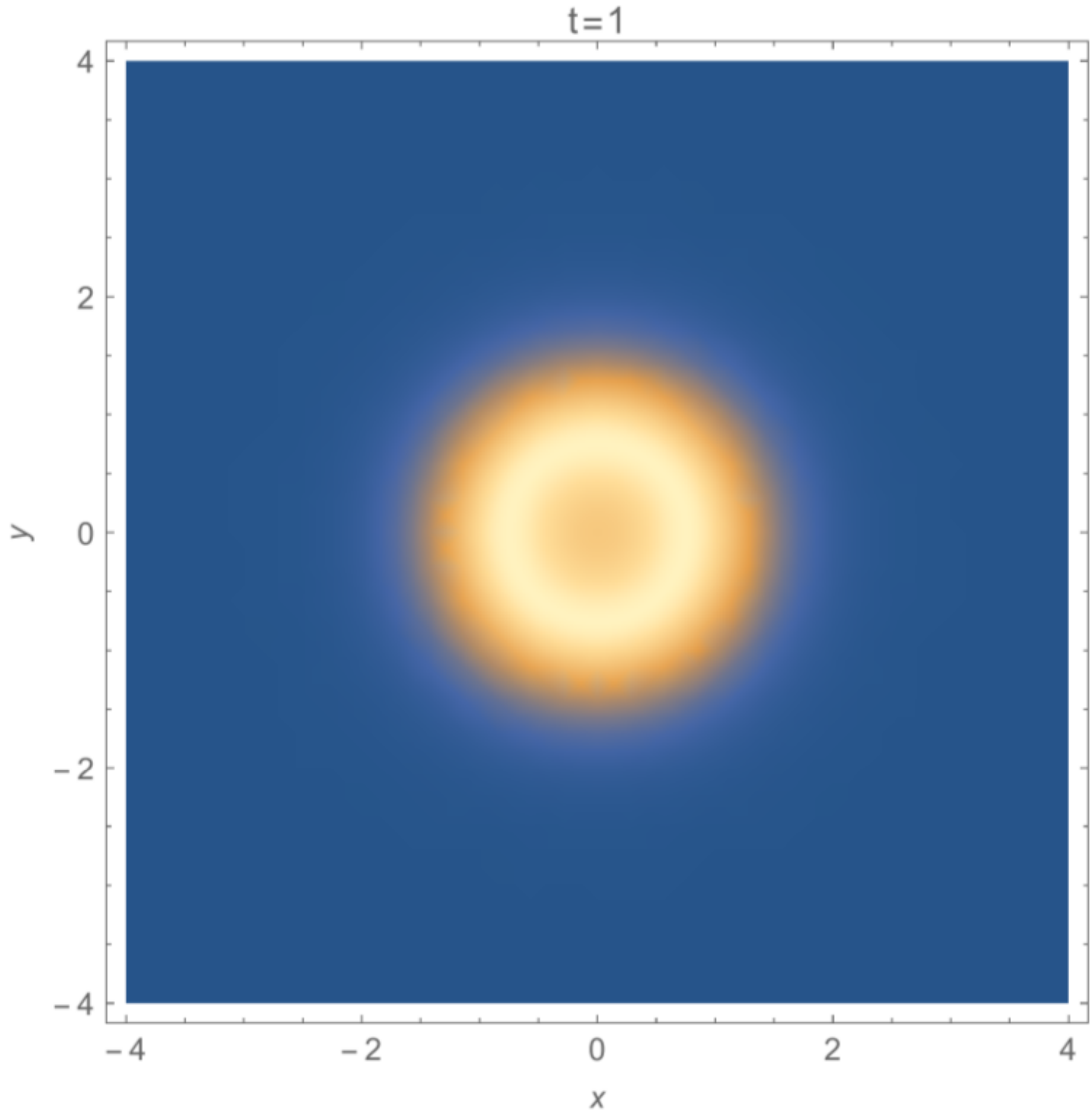}
        \caption{}
        \label{fig:HopfionDPt1}
    \end{subfigure}
    \hfill
    \centering
   \begin{subfigure}[b]{0.2 \textwidth}
        \centering
        \includegraphics[width=\textwidth]{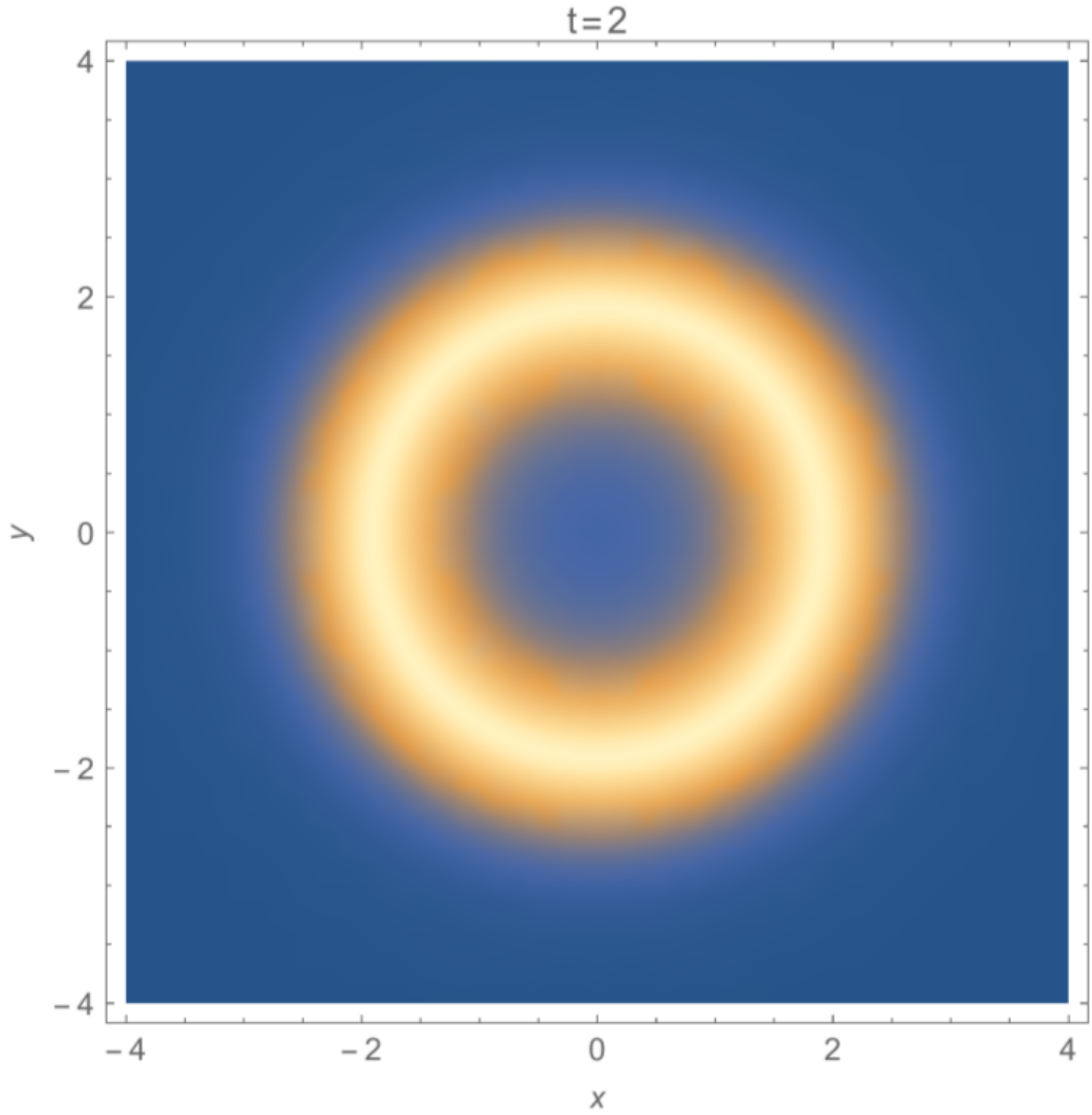}
        \caption{}
        \label{fig:HopfionDPt2}
    \end{subfigure}
    \hfill
    \centering
    \begin{subfigure}[b]{0.2 \textwidth}
        \centering
        \includegraphics[width=\textwidth]{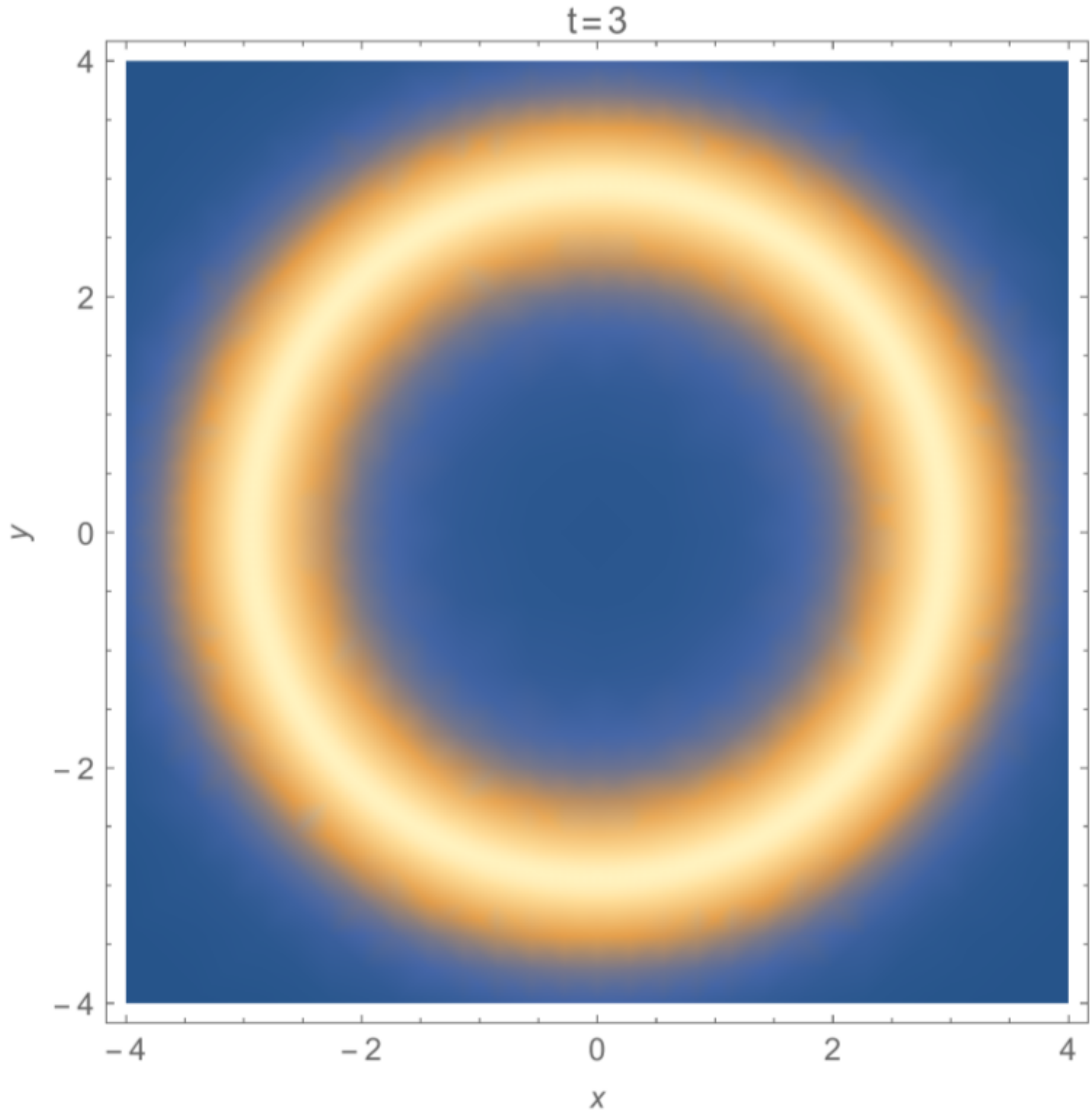}
        \caption{}
        \label{fig:HopfionDPt3}
    \end{subfigure}
    \hfill
    \centering
    \begin{subfigure}[b]{0.2 \textwidth}
        \centering
        \includegraphics[width=\textwidth]{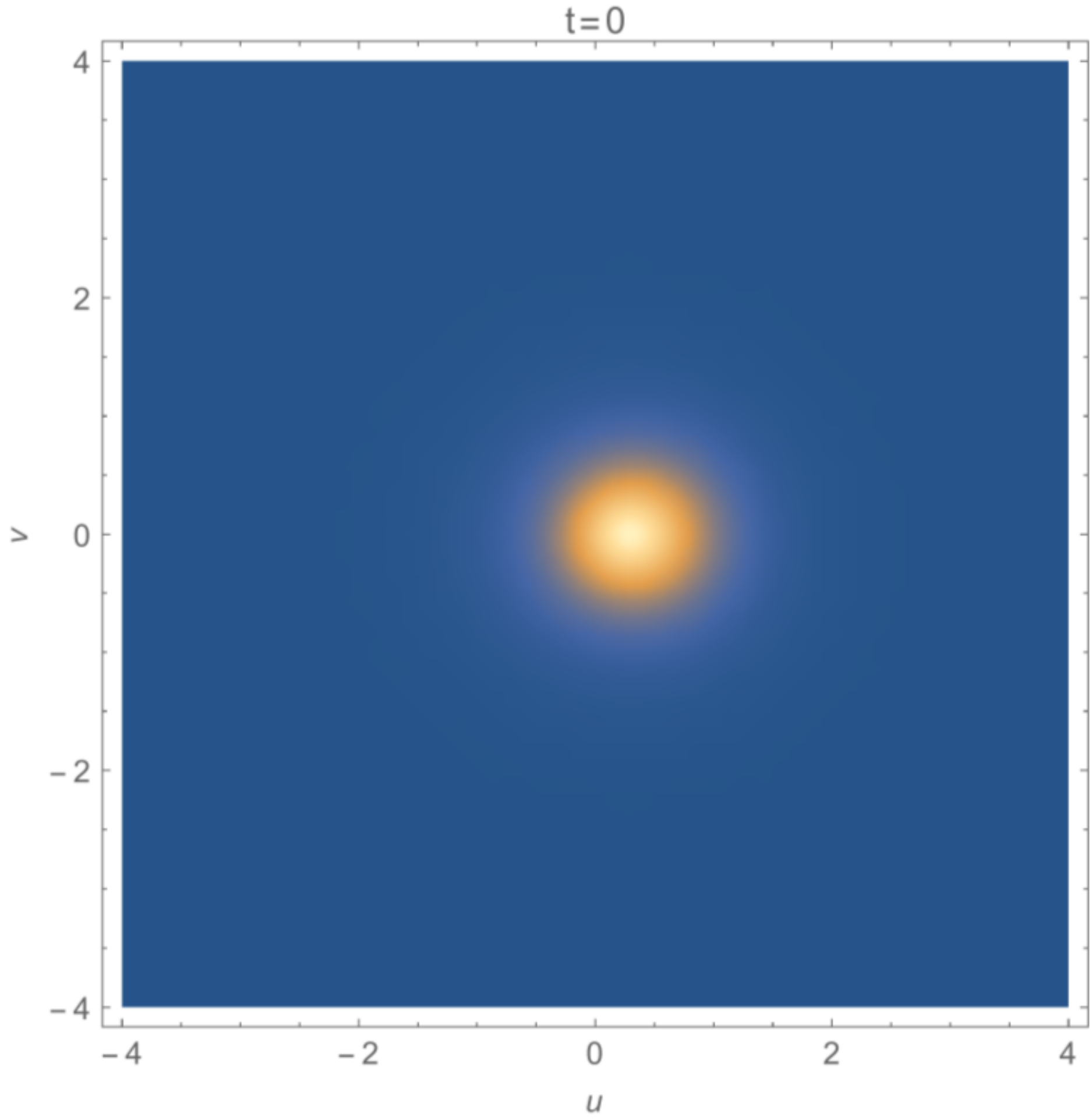}
        \caption{}
        \label{fig:HopfionSCTDPt0}
    \end{subfigure}
    \hfill
    \centering
    \begin{subfigure}[b]{0.2 \textwidth}
        \centering
        \includegraphics[width=\textwidth]{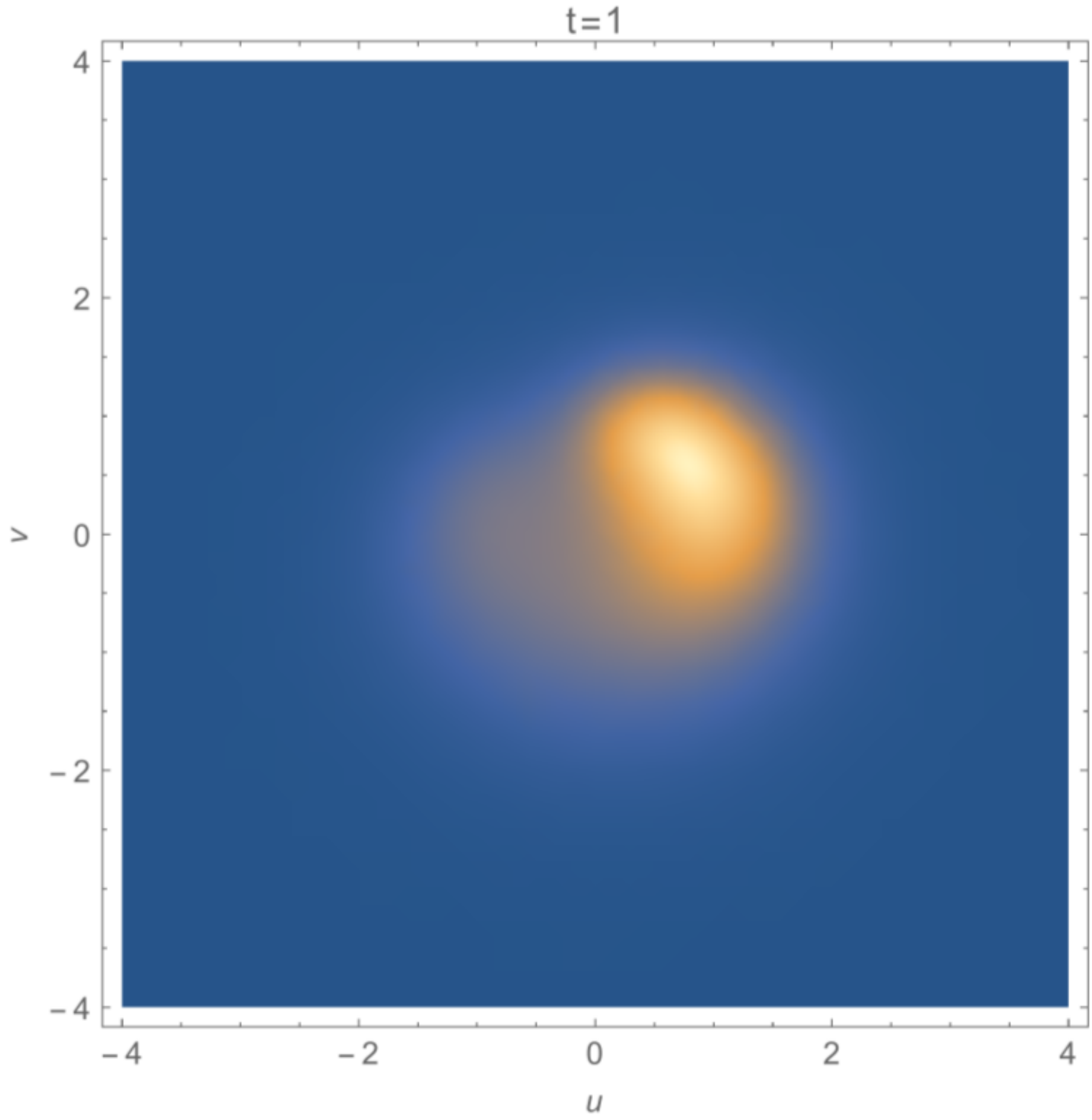}
        \caption{}
        \label{fig:HopfionSCTDPt1}
    \end{subfigure}
     \hfill
    \centering
    \begin{subfigure}[b]{0.2 \textwidth}
        \centering
        \includegraphics[width=\textwidth]{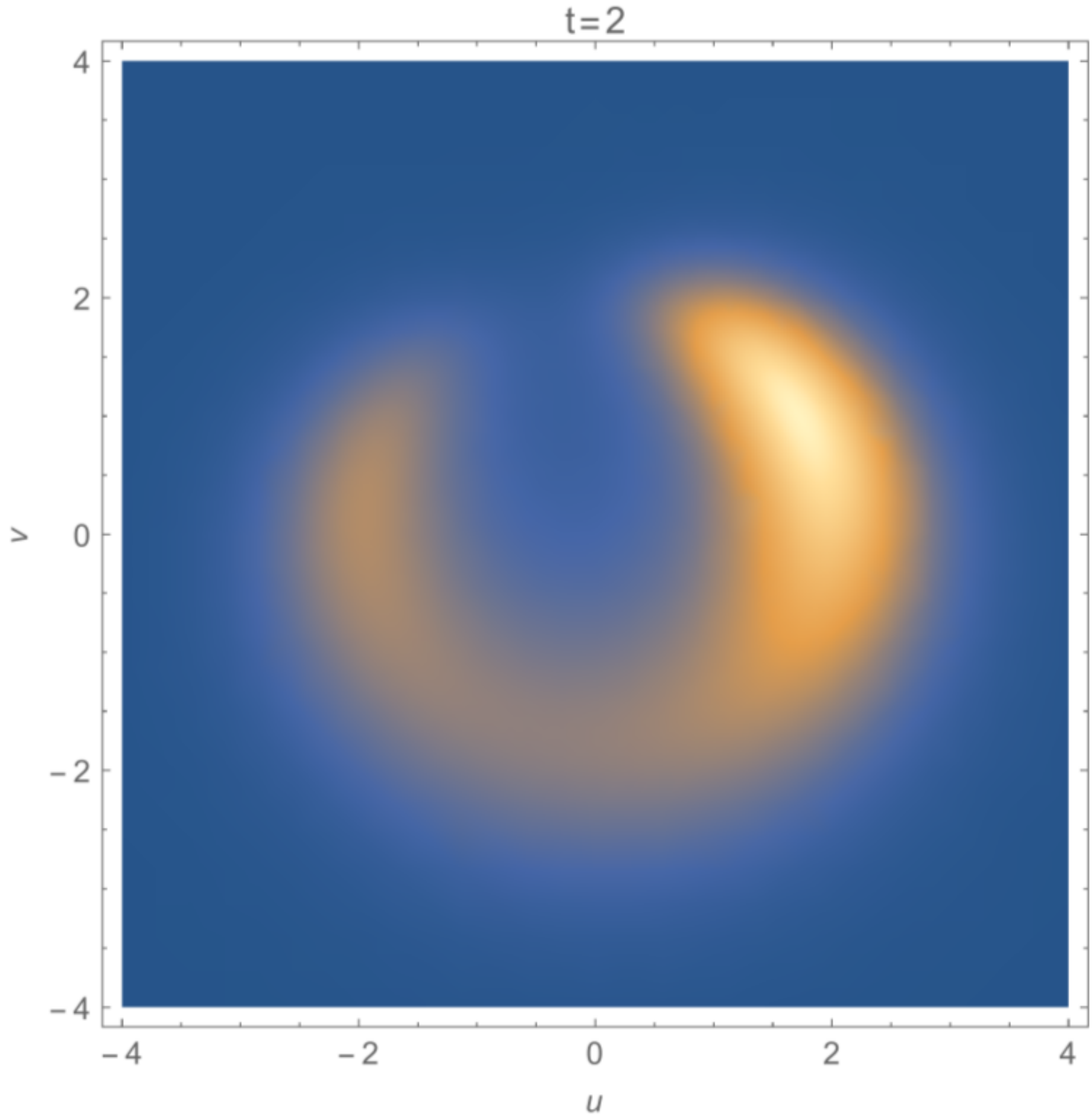}
        \caption{}
        \label{fig:HopfionSCTDPt2}
    \end{subfigure}
     \hfill
    \centering
    \begin{subfigure}[b]{0.2 \textwidth}
        \centering
        \includegraphics[width=\textwidth]{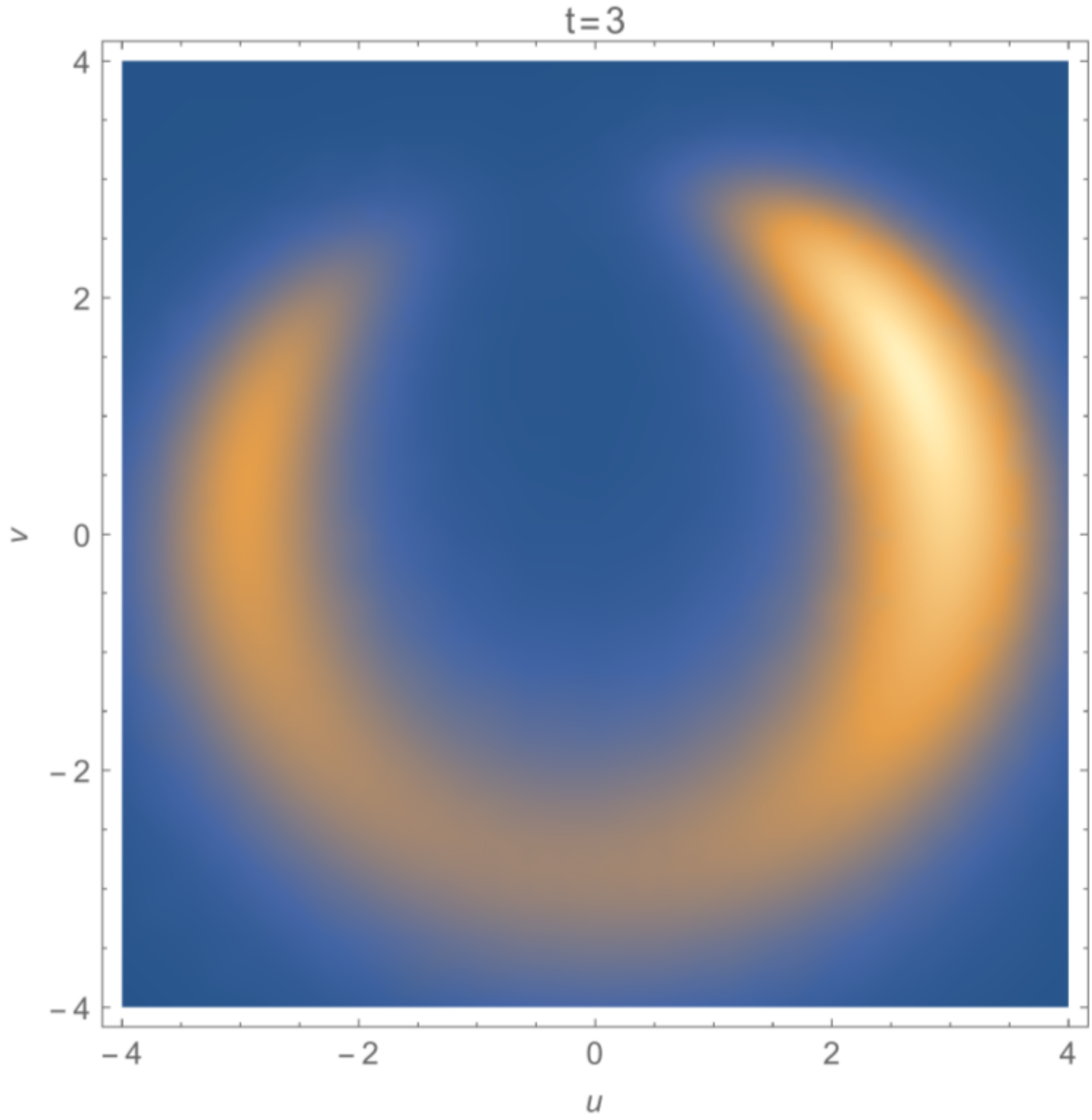}
        \caption{}
        \label{fig:HopfionSCTDPt3}
    \end{subfigure}
    \caption{Plot of the energy density on a plane orthogonal to the total momentum at various times. Time increases
    from left to right.
    Fig(\ref{fig:HopfionDPt0})-Fig(\ref{fig:HopfionDPt3}): Plots for various times for the Hopfion. 
    Fig(\ref{fig:HopfionSCTDPt0})-Fig(\ref{fig:HopfionSCTDPt3}): Plots for various times for the solution found from 
 doing a  SCT of the Hopfion with $b^{\mu}$ as $(0,\imath \half,0,0)$. 
   }
    \label{fig:HopfionSCTDP}
\end{figure}

Let us consider the example where the solution is mapped to a solution with constant electric and magnetic field. For 
$b^{\mu}=\imath(-1,0,0,0)$, the new $\alpha,\beta$ obtained by this map is given by \eqref{abconstantF}.
The corresponding electric and magnetic field are constants as we have seen before, 
and thus have diverging total energy and helicity. We can study
solutions generated by $b^{\mu}=\imath(-b,0,0,0)$, with $b$ approaching unity. 
The radius of the compact electric and magnetic lines  increase monotonically to $\infty$ as $b$ approaches unity. This 
is consistent with the fact that at $b=1$, we have solutions with constant electric and magnetic fields.
\end{itemize}

\section{Formal aspects of Bateman's construction and knot solutions}

In this section we present a series of results regarding Bateman's construction and knot solutions  that we find interesting and that we haven't found explicitly in the previous literature. A summary is the following:
\begin{itemize}
\item We write Bateman's construction in a covariant form.
\item We give expressions for the helicities in terms of integrals over the target space of $\alpha$, $\beta$. We show explicitly that the cross helicities are zero for $(p,q)$-knots.
\item We present a quaternionic formulation of Bateman's construction and relate knot solutions to winding solutions of a non-Abelian gauge theory. 
\end{itemize}

\subsection{Covariant formulation}

Our goal is to write the equations (\ref{absolanz1}) and (\ref{eomab}) in a covariant way\footnote{It was pointed out by an unknown
referee that covariant formulation of Bateman construction was also considered in\cite{VanEnk}.}.
We will be using mostly plus signature for the space-time metric.
The Levi-Civita symbol is defined as,
\be
\epsilon_{ijk}=\epsilon^{ijk}.
\ee
Eq.~(\ref{absolanz1}) can be written as,
\be
E^i + \imath B^i = \epsilon^{ijk} \partial_i \alpha \partial_j \beta.
\ee
Rewriting the equation above in terms of the anti-symmetric field strength ($F_{\mu \nu}$),
\be
F^{0i} + \imath \half \epsilon^{ijk} F_{jk} = \epsilon^{ijk} \partial_i \alpha \partial_j \beta.
\ee
where we have used $E^i=F^{0i}$ and $B^i= \half \epsilon^{ijk} F_{jk}$ .
This can be written in covariant form as,
\be
F^{\mu \nu}- \imath \half \epsilon^{\mu \nu \alpha \beta} F_{\alpha \beta} = -\epsilon^{\mu \nu \rho \sigma} \partial_{\rho} \alpha\partial_{\sigma} \beta.
\label{covarianteq}
\ee
where we define the epsilon tensor as,
\be
\epsilon_{0ijk}=\epsilon_{ijk};~~\epsilon^{0ijk}=-\epsilon^{ijk}.
\ee
Now the $(i,j)$-components of the Eq.~(\ref{covarianteq}) are
\be
F^{ij}-\imath \epsilon^{ijk0} F_{k0} = - \epsilon^{ijk0} (\partial_k \alpha \partial_0 \beta-\partial_0 \alpha \partial_k \beta).
\ee
This is the same as,
\be
\epsilon^{ijk} B_k - \imath \epsilon^{ijk} E_k =  \epsilon^{ijk} (\partial_t \alpha \partial_k \beta-\partial_k \alpha \partial_t \beta).
\ee
which is equivalent to,
\be
 B_k - \imath  E_k =  (\partial_t \alpha \partial_k \beta-\partial_k \alpha \partial_t \beta).
\ee
Multiplying both sides by $\imath$,
\be
E_k + \imath B_k = \imath (\partial_t \alpha \partial_k \beta-\partial_k \alpha \partial_t \beta).
\ee
which is the same as Eq.~(\ref{eomab}). Therefore \eqref{covarianteq} contains all the information about the equations of motion.  If we use the identity,
\be
\epsilon_{\mu\nu\rho_1\sigma_1}\epsilon^{\mu\nu\rho_2\sigma_2 }=-2 \left( \delta^{\rho_2}_{\rho_1} \delta^{\sigma_2}_{\sigma_1}-
\delta^{\sigma_2}_{\rho_1} \delta^{\rho_2}_{\sigma_1}\right),
\ee
we can show that the combination of field strengths appearing in Bateman's construction is imaginary anti-self dual
\be
- \imath \half \epsilon^{\sigma\rho}_{\ \ \mu \nu }  \left[ F^{\mu \nu}- \imath \half \epsilon^{\mu \nu \alpha \beta} F_{\alpha \beta} \right] =F^{\sigma \rho}- \imath \half \epsilon^{\sigma \rho \alpha \beta} F_{\alpha \beta}  .
\ee

We can also obtain covariant formulas for the potentials in terms of $\alpha$ and $\beta$.
From~(\ref{covarianteq}),
\bea
F^{\mu \nu} &=& -\epsilon^{\mu \nu \rho \sigma} \text{Re}\left(\partial_{\rho} \alpha\partial_{\sigma} \beta \right) \label{covarianteqp1},\\
\epsilon^{\mu \nu \alpha \beta} F_{\alpha \beta} &=& 2 \epsilon^{\mu \nu \rho \sigma} \text{Im}\left(\partial_{\rho} \alpha\partial_{\sigma} \beta \right).
\label{covarianteqp2}
\eea
We can rewrite the second equation (\ref{covarianteqp2}) as,
\bea
F_{\mu \nu} &=& \text{Im}\left( \partial_{\mu} \alpha \partial_{\nu} \beta-\partial_{\nu} \alpha \partial_{\mu} \beta\right) \nonumber \\
&=& \partial_{\mu} A_{\nu} -\partial_{\nu} A_{\mu},
\eea
where,
\be
A_{\mu}= \text{Im}\left( H_{\mu} \right)~;~H_{\mu} = \half \left( \alpha \partial_{\mu} \beta -\beta\partial_{\mu}\alpha  \right).\label{gaugepot1}
\ee
The expression of the gauge potential $A_{\mu}$ or the complex potential $H_{\mu}$ is unique up to gauge transformations. We can 
also define a complex field strength corresponding to $H_{\mu}$ as $H_{\mu \nu}=\partial_{\mu}H_{\nu} -\partial_{\nu} H_{\mu}$,
so that $F_{\mu \nu} = \Im \left( H_{\mu \nu}\right)$.

Similarly let us define $C_{\mu}$ and $G_{\mu \nu}$ as,
\be
G_{\mu \nu}=\half \epsilon_{\mu \nu \rho \sigma} F^{\rho \sigma}= \p_{\mu} C_{\nu} - \p_{\nu}C_{\mu}.
\ee
Then, up to a gauge choice,
\be
C_{\mu} = \text{Re} \left(H_{\mu}  \right). \label{gaugepotC1}
\ee
Spatial components of $A_{\mu},C_{\mu}$ and related to electric and magnetic fields as,
\be
E^{i} = \epsilon^{ijk} \p_j C_k ~~;~~B^{i} = \epsilon^{ijk} \p_j A_k.
\ee
Then the first equation (\ref{covarianteqp1}) can be rewritten in terms of the complex field strength as,
\be
\Im (H^{\mu \nu})=- \half \epsilon^{\mu \nu \rho \sigma}  \Re(H_{\rho \sigma})\label{eomcomplex}
\ee
The equation above along with the definition of $H_{\mu \nu}$ in terms of the potential is equivalent to Maxwell's
equations of motion.
We can also define a helicity or Abelian Chern Simons current as,
\be
\kappa^{\mu}= \epsilon^{\mu\nu \rho \sigma} A_{\nu} F_{\rho \sigma} .\label{defk}
\ee
We can show that this current is conserved for the null solutions as,
\be
\p_{\mu} \kappa^{\mu} =\half \epsilon^{\mu\nu \rho \sigma} F_{\mu \nu} F_{\rho \sigma} \propto \mb{E} \cdot \mb{B} =0.\label{AbelianCSCcons}
\ee
also,
\be
\kappa^{0}= -\epsilon^{ijk}A_i F_{jk} =- 2 A_i B^i ~~;~~ h_{mm} = -\half \int d^3 x \kappa^0
\ee
\subsection{Helicities}\label{secAbelianFormII}

In most cases we have to compute the conserved charges of a given solution by evaluating the charge densities on the solution and doing a complicated integral over space. This makes the comparison between different kind of solutions quite cumbersome. However, for the $(p,q)$-knots it turns out that the situation is dramatically improved for the helicities. Indeed, one can find closed expressions in terms of integrals over the target space, which are valid independently of the particular form of the solution in space and time. It is even possible to prove that the cross helicities vanish, as we will do in the following.

Let us define the complex two-form field $H$, related to the electromagnetic field strength $F$ (not to be
confused with RS vector $\mb{F}$) as
\be
H = \imath F +  {}^\star F=df\wedge dg,
\ee
where ${}^\star$ represents Hodge dual, which in terms of the components,
\be
({}^\star F)_{\mu \nu}=\half  \epsilon_{\mu \nu \rho \sigma} F^{\rho \sigma}.
\ee
The relation to the complex vector $\mb{F}=\mb{E}+i\mb{B}$ is
\begin{equation}
H_{i0}=i (\mb{F})_i, \ \ H_{ij}=i\epsilon_{ijk} (\mb{F})^k.
\end{equation}
It will be convenient to solve the field strength and the dual in terms of the complex form $H$,
\bea
F &=& \frac{H-\bar{H}}{2 \imath} \\
{}^\star F &=& \frac{H+\bar{H}}{2}
\eea
where $\bar{ H}$ is the complex conjugate of $H$. 
Let $F=dA$, ${}^\star F = dC$ and $H=dh$. Then, the potentials can be written in the following form
\bea
A &=& \frac{h-\bar{h}}{2 \imath},\\
C &=& \frac{h+\bar{h}}{2},\\
h &=& \half ( f dg - g d f).
\eea
The cross helicity density $h_{em}$ is given by space integral of the following expression,
\be
\chi_{em}= C \wedge F =  \frac{1}{4 \imath}  \left(\bar{h} \wedge H- h \wedge \bar{H} \right). \label{defchiem}
\ee
For $(p,q)$-knot solutions we simply need to evaluate $\chi_{em}$  for $f=\alpha^p, ~g=\beta^q$,
\bea
h &=& \half \left( \alpha^p d(\beta^q) - \beta^q d (\alpha^p) \right)=\alpha^{p-1} \beta^{q-1}\half ( q \alpha d\beta - p \beta d \alpha), \label{relhab1}\\
H &=& d (\alpha^p) \wedge d (\beta^q) = pq \alpha^{p-1} \beta^{q-1} d \alpha \wedge d \beta ,\label{relHab1}\\
|\alpha|^2+|\beta|^2 &=& 1 .\label{abconstraint1}
\eea
Substituting above in Eq.~~(\ref{defchiem}),
\be
\chi_{em}= \frac{pq}{8 \imath} |\alpha|^{2(p-1)} |\beta|^{2(q-1)} \left( q ~ d(\alpha \bar \alpha) \wedge d \beta \wedge d\bar{\beta}+
p ~ d (\beta\bar{\beta} )\wedge d \alpha \wedge d \bar{\alpha} \right) =0.
\ee
Where we have used  Eq.~~(\ref{abconstraint1}) to show that it vanishes. Since $h_{em}=-h_{me}$, both the cross helicities are zero for this class of solutions.

The helicity ($h_{mm}$) is given by space integral of the following form,
\be
\chi_{mm}= A \wedge F = \frac{1}{4}  \left( h \wedge \bar{H} +\bar{h} \wedge H\right).
\ee
Let us try to evaluate $\chi$  again for  solutions of type ($f=\alpha^p, ~g=\beta^q$) where $\alpha,\beta$
satisfies Eq.~(\ref{abconstraint1}). $H,h$ are given by Eq.~(\ref{relHab1}) and (\ref{relhab1}).
We can parametrize $\alpha,\beta$ as,
\bea
\alpha &=& \cos(\phi) e^{\imath \theta_1}, \\
\beta &=& \sin(\phi) e^{\imath \theta_2}, \\
d \alpha &=& -\sin(\phi) e^{\imath \theta_1} d \phi + \imath \cos(\phi) e^{\imath \theta_1} d\theta_1 ,\\
d \beta &=& \cos(\phi) e^{\imath \theta_2} d \phi + \imath \sin(\phi) e^{\imath \theta_2} d\theta_2 .
\eea
From these formulas we can show,
\bea
d \beta \wedge d \alpha \wedge d \bar{\alpha} &=& -2 \sin^2(\phi) \cos(\phi) e^{\imath \theta_2} d\phi \wedge d\theta_1 \wedge d\theta_2,\\
d \alpha \wedge d \beta \wedge d \bar{\beta} &=& -2 \sin(\phi) \cos^2(\phi) e^{\imath \theta_1} d\phi \wedge d\theta_1 \wedge d\theta_2.\\
\eea
Also,
\bea
\chi_{mm} &=& \frac{pq}{8} |\alpha|^{2(p-1)} |\beta|^{2(q-1)} \left( q \bar{\alpha} d\alpha \wedge d \beta \wedge d\bar{\beta}+
p \bar{\beta} d \beta \wedge d \alpha \wedge d \bar{\alpha} + \text{cc}\right)\\
&=& -\frac{pq}{2} \cos^{2p-1} \phi  \sin^{2q-1} \phi \left(q \cos^2\phi + p \sin^2\phi \right) d\phi \wedge d\theta_1 \wedge d\theta_2. \label{ACSdensity}
\eea
Note that $\alpha$ and $\beta$ for the Hopfion solution at time $t=0$ are simply the inverse of the stereographic
projection of the three-sphere $|\alpha|^2+|\beta|^2=1$ onto the space. Since this is a one-to-one mapping, 
we can trade the integral over space for an integral over the $S^3$ parametrized by $\alpha$ and $\beta$.  Then,
\begin{eqnarray}
h_{mm}&=& -\int_{S^3} \chi_{mm} \nonumber \\
& =&  \frac{4 \pi^2 pq}{2}\int_{0}^{\pi/2} d\phi \left(q \cos^{2p+1}\phi \sin^{2q-1}\phi+p\cos^{2p-1} \phi  \sin^{2q+1}\phi\right).
\end{eqnarray}
Now, using the identity
\be
\int_{0}^{\pi/2} d \theta \sin(\theta)^m \cos(\theta)^n = \frac{\Gamma(\frac{m+1}{2})\Gamma(\frac{n+1}{2})}{2 \Gamma(\frac{n+m+2}{2})},
\ee
we find the following value of the helicity
\be
h_{mm} =  2 \pi^2 p q \frac{\Gamma(p+1) \Gamma(q+1)}{\Gamma(p+q+1)} =  2 \pi^2 p q \frac{p! q!}{(p+q)!}.
\ee
for $p=q=1$,$ \int_{S^3} \chi =  \pi^2$. Note the result is symmetric under interchange of $(p,q)$.
For more general holomorphic functions of $\alpha,\beta$ satisfying Eq.~(\ref{abconstraint1}), we find the following expressions for the helicity density :
\bea
h &=& \half \left( f(\alpha,\beta) d(g(\alpha,\beta)) - g(\alpha,\beta d (f(\alpha,\beta) \right), \\
H &=& d (f(\alpha,\beta)) \wedge d (g(\alpha,\beta),\\
\chi_{mm} &=& \half \Big\lbrack \Re\left( \frac{(\p_{\alpha} f \p_{\beta} g-(\p_{\alpha} g \p_{\beta} f)(f \p_{\alpha} g-g \p_{\alpha}f)}{\beta}\right) \sin^3(\phi) \cos(\phi) \nonumber \\
&-& \Re\left( \frac{(\p_{\alpha} f \p_{\beta} g-(\p_{\alpha} g \p_{\beta} f)(f \p_{\beta} g-g \p_{\beta}f)}{\alpha}\right) \sin(\phi) \cos^3(\phi) \Big\rbrack \nonumber \\
&& d\phi \wedge d\theta_1 \wedge d\theta_2 .\label{ACSdensityg}
\eea

\subsection{Quaternionic Formulation and Map to Non Abelian theories}

There is an interesting map between Bateman's construction and quaternions that can be used to embed the Hopfion solution 
in a non-Abelian Yang-Mills theory. The non-Abelian solution is pure gauge with winding number different than zero. 
Similar embeddings of knot solutions  were studied in the past \cite{Jackiw:1999bd,vanBaal:2001jm,Cho:2004bz}, 
the main differences in this case are that the knot is not static and that it is a solution of Maxwell's theory.

Using $\alpha$ and $\beta$ we can define a quaternion $\mb{q}$,
\bea
\mb{q} &=& \frac{1}{m} (\alpha + \beta j), \label{defq} \\
m &=& \sqrt{|\alpha|^2+|\beta|^2}, \\
\alpha &=& \alpha_0 + \alpha_1 i ~~;~~\beta = \beta_0 + \beta_1 i, 
\eea
where $\alpha_0,\alpha_1, \beta_0,\beta_1$ are real functions depending on space-time.
The conjugate quaternion is
\be
\mb{q}^*=\frac{1}{m}(\alpha^* - \beta j) .\label{defqs}
\ee
We will define the following quaternionic-valued potential:
\bea
\mb{Q}_{\mu} &=&\mb{q} (\partial_{\mu} \mb{q}^*)\label{defQd}.
\eea
We can easily show that,
\be
\mb{Q}_{\mu} = -\mb{Q}_{\mu}^*
\ee

We will write the components of the quaternionic potential in the following form:
\be
\mb{Q}_{\mu} =   \frac{1}{m^2} \Im (\alpha \p_{\mu} \alpha^*+ \beta \p_{\mu} \beta^*) i - \frac{2}{m^2} H_{\mu} j.
\ee
where the complex potential $H_{\mu}$ is defined in \eqref{gaugepot1}.

Associated to the quaternionic potential we can define a quaternionic field strength,
\be
\label{defQdd}
\mb{Q}_{\mu \nu} = \partial_{\mu}  \mb{Q}_{\nu}-\partial_{\nu}  \mb{Q}_{\mu} = \left( \partial_{\mu} \mb{q} \partial_{\nu} \mb{q}^*-\partial_{\nu} \mb{q} \partial_{\mu} \mb{q}^* \right).
\ee
Then,
\bea
\mb{Q}_{\mu \nu} &=&  \frac{2}{m^2} \Im\left( \p_{\mu} \alpha \p_{\nu} \alpha^* + \p_{\mu} \beta \p_{\nu} \beta^*\right) i-\frac{2}{m^2} H_{\mu \nu}j + \mb{J}_{\mu \nu} \label{expQmunu}\\
\mb{J}_{\mu \nu} &=& m^2\partial_{\mu} (\frac{1}{m^2}) \mb{Q}_{\nu} - m^2\partial_{\nu} (\frac{1}{m^2}) \mb{Q}_{\mu}\label{defJmunu}
\eea
Note that $\mb{J}_{\mu \nu}=0$ for cases where $m$ is constant.
The equation of motion given by~(\ref{eomcomplex}) can be rewritten as,
\be
(\mb{Q}^{\mu \nu} -\mb{J}^{\mu \nu})_k= - \half \epsilon^{\mu \nu \rho \sigma} (\mb{Q}_{\rho \sigma} -\mb{J}_{\rho \sigma})_j,
\ee
Since,
\be
\mb{q} \mb{q}^*=1 \label{constraint1}.
\ee
Then we can show,
\bea
\mb{Q}_{\mu \nu} &=& \partial_{\mu}  \mb{Q}_{\nu}-\partial_{\nu}  \mb{Q}_{\mu} = - \left[  \mb{Q}_{\mu},  \mb{Q}_{\nu} \right].
\label{constraint2}
\eea

Note that we can interpret \eqref{defQd} as a non-Abelian $SU(2)$ gauge potential
by using Pauli matrices as a representation of the quaternion components $1,i,j,k$
\be
1 \equiv \mb{1}_{2\times 2},~~i\equiv -i \sigma_1,~~j\equiv -i \sigma_2,~~k \equiv -i \sigma_3.
\ee
Using \eqref{constraint2} it is clear that the non-Abelian field strength is zero, and therefore this is a pure gauge solution. We will show now that the winding number is non-zero, and in fact it is directly related to the helicities of the Abelian solution.

\subsubsection{Winding number of non-Abelian solutions}

The winding number of  non-Abelian solutions is measured by the integral over space of the Chern-Simons three-form
\begin{equation}
w=-\frac{1}{8\pi^2} \int_{S^3}\, {\rm tr}\,\left(\mb{ K}\right),
\end{equation}
where the Chern-Simons three form is
\begin{equation}
\mb{ K}=\mb{ Q}\wedge d\mb{ Q}+\frac{2}{3}\mb{ Q}\wedge\mb{ Q}\wedge\mb{ Q}.
\end{equation}
For smooth pure gauge non-Abelian configurations in a compact space $w$ is an integer. 
We can define a current one-form using the Hodge dual $\mb{J}_{CS}={}^\star\mb{ K}$. We can show that the current is conserved when the non-Abelian pure gauge configuration is obtained from a null Abelian solution  obeying Eq.~(\ref{constraint2}) ($d\mb{Q}=-\mb{Q}\wedge \mb{Q}$)
\begin{eqnarray}
{}^\star d {}^\star \mb{J}_{CS} &=& {}^\star d\mb{K}= {}^\star(d\mb{Q}+\mb{Q}\wedge \mb{Q})\wedge (d\mb{Q}+\mb{Q}\wedge \mb{Q})-{}^\star(\mb{Q}\wedge\mb{Q}\wedge\mb{Q}\wedge\mb{Q}) \nonumber\\
&=&-{}^\star(d\mb{Q}\wedge d\mb{Q})\propto \mb{E}\cdot \mb{B}=0.\label{non-AbelianCSCcons}
\end{eqnarray}
The non-Abelian Chern-Simons  current for a solution obeying Eq.~(\ref{constraint2}) is 
\begin{equation}
\mb{J}_{CS}=\frac{1}{3}{}^\star(\mb{Q}\wedge d\mb{Q}).
\end{equation}
In components, this is
\be
{\mb{J}_{CS}}^{\mu} =\frac{1}{6} \epsilon^{\mu \nu \rho \sigma} \mb{Q}_{\nu} \mb{Q}_{\rho \sigma} =\frac{1}{3}  \epsilon^{\mu \nu \rho \sigma} \mb{Q}_{\nu}\mb{Q}_{\rho}\mb{Q}_{\sigma}.
\ee
Using $\mb{Q}_{\mu}^*=-\mb{Q}_{\mu}$ we can show,
\be
{\mb{J}_{CS}}^{\mu} = ({\mb{J}_{CS}}^{\mu})^*.
\ee
Let us evaluate ${\mb{J}_{CS}}^{\mu}$ in terms of components,
\be
{\mb{J}_{CS}}^{\mu} = -\frac{1}{6 } \epsilon^{\mu \nu \rho \sigma} \sum_{a \in \{i,j,k\}}(\mb{Q}_{\nu})_a  (\mb{ Q}_{\rho \sigma})_a+ \cdots
\ee
The $\cdots$ denotes terms that are proportional to $i,j,k$, which we can set to zero even without calculation as 
${\mb{J}_{CS}}^{\mu}$ is real. Now using \eqref{defJmunu},
\be
{\mb{J}_{CS}}^{\mu} = -\frac{1}{6 } \epsilon^{\mu \nu \rho \sigma} \sum_{a \in \{i,j,k\}}(\mb{Q}_{\nu})_a  (\mb{ Q}_{\rho \sigma}-\mb{J}_{\rho \sigma})_a
\ee
We can show (there is no summation over the index $a$),
\be
\epsilon^{\mu \nu \rho \sigma} (\mb{ Q}_{\nu})_a  (\mb{ Q}_{\rho \sigma}-\mb{J}_{\rho \sigma})_a 
=\frac{4}{m^2}  \epsilon^{\mu \nu \rho \sigma} A_{\nu} F_{\rho \sigma} \mb{1}_{2\times 2}= \frac{4}{m^2} \kappa^{\mu} \mb{1}_{2\times 2}~\forall a \in (i,j,k),
\ee
where $\kappa^{\mu}$ \eqref{defk} is the Abelian Chern Simons current. So the relation between Abelian and non-Abelian (quaternionic) Chern Simons
is given by,
\be
{\mb{J}_{CS}}^{\mu}= \frac{2}{m^4} \kappa^{\mu} \mb{1}_{2\times 2}.
\ee
Both of the currents, ${\mb{J}_{CS}}^{\mu}$ and $\kappa^{\mu}$ were shown to be  conserved in 
\eqref{non-AbelianCSCcons} and \eqref{AbelianCSCcons}.
This gives a non-trivial constraint on the derivative of $m$ : $\kappa^{\mu}\p_{\mu}m=0$.

For $\alpha=\tilde \alpha^p,~\beta=\tilde \beta^q$ with $|\tilde \alpha|^2+|\tilde \beta|^2=1$,
$\int d^3 x \kappa^0=2\int \chi_{mm}$ is calculated for these solutions in Eq.~(\ref{ACSdensity}). Then the
corresponding winding number of the non-Abelian solution is given by
\begin{eqnarray}
w &=& -\frac{1}{8 \pi^2} \int_{S^3} \tr {\mb{J}_{CS}}^{0} = -2 \frac{1}{8 \pi^2}  \int_{S^3} \frac{4}{m^4} \chi_{mm}  \nonumber \\
 &=&    2 \frac{1}{8 \pi^2} \int_{0}^{\pi/2} d\phi \frac{4}{(\cos(\phi)^{2p}+\sin(\phi)^{2q})^2}\frac{4 \pi^2 pq}{2}\left(q \cos^{2p+1}\phi \sin^{2q-1}\phi+p\cos^{2p-1} \phi  \sin^{2q+1}\phi\right) \nonumber \\
 &=& p q.
\end{eqnarray}

In contrast with the values of the Abelian helicities, which were not integers, the corresponding non-Abelian winding number is 
quantized in integer values. Note that the Hopfion solution has winding number $w=1$ as expected. 
We can also use the Abelian Hopfion solution to construct higher winding non-Abelian solutions.
Consider the non- Abelian (quaternionic)  potential,
\be
\mb{\tilde  Q}_{\mu} =   \mb{\tilde q}(\partial_{\mu} \mb{\tilde q}^*)
\ee
where,
\be
\mb{\tilde q}= \tilde \alpha + \tilde \beta j = q^n = (\alpha + \beta j)^n
\ee
Note,
\be
\mb{\tilde q} \mb{\tilde q}^* = (\mb{q} \mb{q}^*)^n=1
\ee
and $\alpha,\beta$ are given by the Abelian solution Eq.~(\ref{hopfalpha1}) and Eq.~(\ref{hopfbeta1}). The non-Abelian
gauge potential $\mb{\tilde Q}_{\mu}$ naturally gives a flat gauge connection due to its structure. We expect these to be
solutions with winding number $n$-times that of the solution with potential given by $\mb{q}$, 
we have checked for $n=2$ that it is indeed the case. These higher winding non-Abelian solutions $\tilde\alpha$ and $\tilde\beta$ are not solutions of Maxwell's equations.

\section{Summary}
In this note we have further developed the study of topologically non-trivial solutions of electrodynamics. We have discovered a novel method of generating such solutions  by applying conformal transformations with complex parameters  on known solutions expressed in terms of Bateman's variables.
This enabled us in fact to get a wide class of  solutions  from the basic configuration of constant electric and magnetic fields. We have introduced a covariant formulation of the Bateman's construction and discussed the conserved charges associated with the conformal group as well as a set of four types of conserved helicities. One way to implement the covariant formulation is to use a quaternionic formulation. This led to a simple map between the electromagnetic knotted solutions into flat connections of $SU(2)$ gauge theory. We computed the corresponding CS charge and show that it takes an integer value.

There is an ample variety of open questions related to the study of the knotted solutions of electrodynamics. Here we mention few of them.
\begin{itemize}
\item
Obviously the most interesting issue is how to realize the Hopfion or any of its cousins  in the Laboratory. 
The question is whether we can supply our experimental colleagues with additional information that will enable them to 
produce these non-trivial electromagnetic configurations. A proposal for constructing Hopf solutions in laboratory was
given in~\cite{IrvineBouwmeester2008}.
\item
The solutions discussed in this note, from the constant $\mb{E}$ and $\mb{B}$ and all the way to the generalization of the 
Hopfions were expressed in Cartesian coordinates. Obviously there are electromagnetic configurations that are naturally described in cylindrical and spherical coordinates.  
One can search for their expressions in terms of the Bateman's variables which will now be complex functions of spherical 
or cylindrical or more general coordinates. Performing the conformal transformation with complex coordinates can 
also be done using various different coordinate systems.
\item
An interesting question is obviously deciphering the trajectories of charged particles in the
topologically non-trivial EM fields. Do these trajectories admit some topology? A class of charged particle
trajectories in Hopfion background was studied numerically in~\cite{ArrayasTrueba}. 
A more general question is looking for similar topological solutions to the equations of motion of electrodynamics coupled to currents and charge densities.
\item
A more theoretical issue is the classification of all solutions that follow from Bateman's construction. The class of solutions based on taking any holomorphic function of the basic Hopfion $\alpha$ and $\beta$, as well as all other solutions that one gets by applying all possible conformal transformations with complex parameters.
\item
A similar issue relates to the classification of the  corresponding $SU(2)$ flat gauge connections. Are there $SU(2)$ flat gauge connections that cannot be mapped into the EM topologically non-trivial configurations? and if yes what characterize those that can be mapped.
\item
We have mentioned that the solutions found in this note are in fact valid in any conformal flat background of which $AdS_4$ (Anti-de Sitter $4$-d space-time) 
is a special case.
One may wonder in the context of holography\cite{Maldacena:1997re},\cite{Gubser:1998bc},\cite{Witten:1998qj}
with electromagnetic fields in the bulk what are the corresponding dual (topological?) configurations on the 
boundary conformal field theory.
\end{itemize}
\section{Acknowledgments}
J.S would like to thank Ori Ganor for useful discussions. We would like to acknowledge the useful comments and suggestion from
unknown referees. 
The work of N.S was supported by "The PBC program for fellowships for outstanding post-doctoral researcher from China and 
India of the Israel council of higher education''. 
This work was partially supported   by the Israel Science Foundation (g
rant 1989/14),the US-Israel bi-national fund (BSF) grant number 2012383 and the Germany–Israel bi-national fund GIF grant number I-244-303.7-2013.
This work is partially supported by the Spanish grant MINECO-13-FPA2012-35043-C02-02. 
C.H. is supported by the Ramon y Cajal fellowship RYC-2012-10370.

\appendix
\section{Noether Charges} ~\label{appendixCharges}

Let us consider a gauge potential $A_{\mu}$ and the corresponding field strength 
$F_{\mu \nu}= \p_{\mu} A_{\nu}-\p_{\nu} A_{\mu}$ to be solutions of $4$-d vacuum Maxwell equations $\p_{\mu} F^{\mu \nu}=0$.
The electric and magnetic field are given by,
\bea
E^i &=& F^{0i}=F_{i0}\\
B^i &=& \half \epsilon^{ijk} F_{jk} ~~\text{or} ~~ F_{jk} =\epsilon_{ijk}  B^i 
\eea
where we have assumed mostly positive signature for the space-time metric.
Then,
\be
-\frac{1}{4} F_{\mu \nu} F^{\mu \nu}=\half (\mathbf{E} \cdot \mathbf{E}-\mathbf{B} \cdot \mathbf{B})
\ee
The stress-energy tensor is
\be
T_{\mu \nu} = -F_{\mu \rho} F^{\rho}_{~ \nu}+\frac{1}{4} \eta_{\mu \nu} F_{\alpha \beta} F^{\alpha \beta}
\ee
Explicit expressions of the components of stress energy tensor are
\bea
T_{00} &=& \half (\mathbf{E} \cdot \mathbf{E}+\mathbf{B} \cdot \mathbf{B}) \\
T_{0i} &=& -(\mathbf{E} \times \mathbf{B})_i\\
T_{ij} &=&  \delta_{ij} \half (\mathbf{E}\cdot\mathbf{E}+\mathbf{B}\cdot\mathbf{B}) - E_i E_j -B_i B_j
\eea
The stress energy tensor satisfies,
\be
\p_{\mu} T^{\mu \nu}=0 ~~~;~~~ T^{\mu}_{\mu}=0
\ee
Then the list of conserved charge densities from conservation of stress tensor are given by,
\bea
&& \text{Energy: }~ \mathcal{E}=T^{00} = \half (\mathbf{E} \cdot \mathbf{E}+\mathbf{B} \cdot \mathbf{B}) \\
&& \text{Momentum :} ~ p^i=T^{0i}=(\mathbf{E} \times \mathbf{B})^i
\eea
Now let us consider the conserved current corresponding to the Lorentz symmetry,
\bea
J^{\mu (\alpha \beta)}_{\Lambda} &=& T^{\mu \alpha} x^{\beta} - T^{\mu \beta} x^{\alpha} \\
\p_{\mu} J^{\mu (\alpha \beta)}_{\Lambda} &=& T^{\beta \alpha} - T^{\alpha \beta} =0
\eea
using the symmetry properties of the stress tensor. The corresponding 
charges are given by, $J^{0 (\alpha \beta)}_{\Lambda}$, which is antisymmetric in $(\alpha,\beta)$.
Then the spatial and temporal charges given by $J^{0 (\alpha \beta)}_{\Lambda}$ is,
\bea
&& \text{Momentum :} ~l^{i} = \sum_{jk} \epsilon^{i j k} J^{0 (j k)}_{\Lambda} = (p \times x)^{i}\\
&& \text{Boost :}  ~ b^{i} = J^{0 (0 i)}_{\Lambda} = \mathcal{E} x^i - p^i t
\eea
Now consider the current corresponding to dilatation symmetry,
\bea
J^{\mu}_{D} &=& T^{\mu}_{\nu} x^{\nu}\\
\p_{\mu} J^{\mu}_{D} &=& T^{\mu}_{\mu} =0
\eea
The corresponding conserved charge density is given by,
\be
\text{Dilatation : }~ d =   J^{0}_{D} = - \mathcal{E} t + \sum_i p^i x^i
\ee
Now let us consider the current corresponding to the special conformal transformation,
\bea
J_{k}^{\mu (\nu)} &=& x^{\alpha} x_{\alpha} T^{\mu \nu} - 2 T^{\mu}_{\rho} x^{\rho} x^{\nu} \\
\p_{\mu} J_{k}^{\mu (\nu)} &=& -2 T^{\mu}_{\mu} x^{\nu}=0
\eea
Then the corresponding  temporal and spatial charges are given by,
\bea
&& \text{TSCT: } ~k^0 = J_{k}^{\mu (0)} = ( x^i x_i+t^2) \mathcal{E} -2 t p_i x^i \\
&& \text{SSCT: } ~ k^i = - J_{k}^{\mu (i)} = 2 x^i p_j x^j-2 \mathcal{E} t x^i-(x^i x_i -t^2)p^i
\eea

\section{$\theta,\phi$ Formulation}

\subsection{Hopf index in $3$-dimensions}

Consider a complex scalar field $\phi(x_1,x_2,x_3)$ defined on $\mathbb{R}^3$. Let the field be well defined as $r \to \infty$,
where $r=\sqrt{x_1^2+x_2^2+x_3^2}$, \ie the map  does not depend on the direction. Then the map can be identified
as $\phi : S^3 \to S^2$, as $S^3 \equiv \mathbb{R}^3 \cup \{ \infty \}$ and $S^2 \equiv \mathbb{C}^1 \cup \{ \infty \}$. These class of maps are
generally classified in various homotopy classes specified by a topological invariant called Hopf index.

Consider the area two form~\cite{Ranada:1992hw},
\be
\mathcal{F} = \frac{1}{2 \pi \imath} \frac{d \phi^* \wedge d \phi}{(1+ \phi \phi^*)^2}
\ee
Since $\mathcal{F}$ is closed in $S^3$, it must also be exact, \ie there is a one form $\mathcal{A}$ such that
$\mathcal{F} = d \mathcal{A}$. Then the Hopf index is given by~\cite{Whitehead},
\be
n = \int_{S^3} \mathcal{A} \wedge \mathcal{F}\label{defHI}
\ee
The pull back of the form $\mathcal{F}$ in $\mathbb{R}^3$ is given by,
\be
\mathcal{F} = \half F_{ij} dx^i \wedge dx^j = \frac{1}{4\pi \imath} \frac{\p_i \phi^* \p_j \phi- \p_j \phi^* \p_i \phi}{(1+ \phi \phi^*)^2} dx^i \wedge dx^j
\ee
We can also express $\mathcal{F}$ in terms of a vector field defined as,
\be
B^i=B_i = \half \epsilon^{ijk} F_{jk}
\ee
$\mb{B}=B_i dx^i=W(\phi)$ is called the Whitehead vector of the map $\phi$. By definition, $\mb{B}\cdot\nabla \phi=0$,
\ie the vector $\mb{B}$ is always tangential to the surfaces $\phi = constant$.

Let us parametrize the map $\phi$ as,
\be
\phi = S e^{2\pi \imath \psi} ~~;~~ \rho = -\frac{1}{1+S^2}
\ee
where $S, \psi,\rho$ are real maps from $\mathbb{R}^3 \to \mathbb{R}$.
then,
\be
\mathcal{F} = d\rho \wedge d \psi ~~;~~ \mathcal{A} = \rho d \psi + d \xi
\ee
$\xi$ is some arbitrary function from $\mathbb{R}^3 \to \mathbb{R}$. Then,
\be
n= \int_{S^3} d(\xi d\rho \wedge d \psi)
\ee
is zero by Gauss' theorem unless $\psi$ or $\rho$ is multiple valued. We will assume $\psi$ to be multiple valued, and $\nabla \psi$
to be a single valued function. 
\paragraph{Hopf Map}
A simplest example of a map $\phi: S^3 \to S^2$ with non zero Hopf index is
\be
\phi_H(x,y,z) = \frac{2(x+ \imath y)}{2 z + \imath (r^2-1)}~~;~~ r^2 =x^2+y^2+z^2 \label{Hopfmap}
\ee
The Hopf index is unity for this map.

\subsection{$(n,m)$-link solutions in $3+1$-dimension}
Now consider two such maps $\phi, \theta : S^3 \times \mathbb{R} \to S^2$. Then let us consider the electric and magnetic field
to be Whitehead vectors of the two maps, $\mb{B} = W(\phi)$ and $\mb{E} = W(\theta)$. The maps now depend on a extra time parameter, and to make the equations consistent with the $3+1$-d Maxwell equations,
\bea
M_{\mu \nu} &=& \half \epsilon_{\mu \nu \rho \sigma} F^{\rho \sigma} \label{relFM}\\
F_{\mu \nu} &=& f_{\mu \nu}(\phi) = \frac{1}{2\pi \imath} \frac{\p_{\mu} \phi^* \p_{\nu} \phi- \p_{\nu} \phi^* \p_{\mu} \phi}{(1+ \phi \phi^*)^2}\label{defF} \\
M_{\mu \nu} &=& f_{\mu \nu}(\theta) = \frac{1}{2\pi \imath} \frac{\p_{\mu} \theta^* \p_{\nu} \theta- \p_{\nu} \theta^* \p_{\mu} \theta}{(1+ \theta \theta^*)^2}\label{defM}
\eea
where~\footnote{we define $\epsilon_{0ijk}=-\epsilon^{0ijk}=\epsilon^{ijk}=\epsilon_{ijk}$},
\be
E_i = F_{i0} = \half \epsilon_{ijk} M^{jk} ~~;~~ B_i = M_{0i} = \half \epsilon_{ijk} F^{jk}
\ee
Note our definitions are the same as that of ~\cite{Ranada:1992hw} up to some signs. For solutions parametrized as (\ref{relFM})-(\ref{defM}), we can show,
\be
\mb{E} \cdot \mb{B} =\half \epsilon_{ijk} M_{0i} M^{jk} =- \half \epsilon^{\mu\nu \rho \sigma}M_{\mu \nu}M_{\rho \sigma}=0
\ee
as a consequence the helicities defined ($\mb{B} = \nabla \times \mb{A}$ and $\mb{E} = \nabla \times \mb{C}$) as,
\bea
h_e &=& \int_{S^3} A \cdot B \label{defhe}\\
h_m &=& \int_{S^3} C \cdot E \label{defhm}
\eea
are time independent. 
 Also these are proportional to the Hopf Index (\ref{defHI}) of the two maps $\theta$ and $\phi$. Since Hopf Indices
are integers, the helicities will be given by integer multiple of some constant number.

\paragraph{Hopfion I}\label{secHopfionI}
The simplest solution with $h_e=h_m$ is given by~\cite{Ranada:1990},\cite{Ranada:1995},\cite{IrvineBouwmeester2008},
\bea
\phi &=& \frac{(A  z +  t (A-1))+ \imath ( t x- A  y)}{(A x +  t y)+ \imath (A (A-1) -t z)} \label{defphi}\\
\theta &=&  \frac{(A  x +  t y)+ \imath (A z + t(A-1))}{( t x - A y)+ \imath (A (A-1) -t z)}\label{deftheta}\\
A&=& \half(x^2+y^2+z^2-t^2+1)
\eea
which at $t=0$ reduces to Hopf map (\ref{Hopfmap}),
\bea
\phi(t=0,x,y,z) &=& \frac{2(z-\imath y)}{2 x + \imath (r^2-1)} = \phi_H(z,-y,x)\\
\theta(t=0,x,y,z) &=& \frac{2(x+\imath z)}{-2 y + \imath (r^2-1)}=\phi_H(x,z,-y)
\eea
Vacuum electrodynamics have the conformal group as a symmetry group.
The list conserved charges and their values are given ~\cite{IrvineBouwmeester2008}(in units in which energy =1) in
table (\ref{table1}). The energy density of the solution given by,
\be
\mathcal{E}=\half (E^2+B^2)=\frac{16 \left( 1+x^2+y^2+(t+z)^2\right)^2}{\pi^2 \left( (1-t^2+x^2+y^2+z^2)^2+4 t^2 \right)^3}
\ee

\begin{table}[h]
\begin{center}
 \begin{tabular}{|c|c|c|}
 \hline
  Charge & Density & Value \\
  \hline
Energy & $\mathcal{E} = \half (E^2+B^2) $ & $1$\\
\hline
Momentum & $ \mb{p} = (\mb{E} \times \mb{B})$ & $(0,0,\half)$\\
\hline
Angular Momentum & $\mb{L} =(\mb{p} \times \mb{x})$ & $(0,0,-\half)$ \\
\hline
Boost vector & $\mb{b}_v = (\mathcal{E} \mb{x} - \mb{p} t)$ &  $(0,0,0)$\\
\hline
SCT & $k^0 = \left((x^2+t^2) \mathcal{E}+ t \mb{p}\cdot \mb{x}\right)$ & $1$\\
\hline
SCT & $\mb{k} = \left(2 \mb{x} (\mb{p}\cdot \mb{x})-2 t \mathcal{E} \mb{x}-x^2 \mb{p} \right)$ & $(0,0,\half)$ \\
\hline
Dilatation &  $d= (\mb{p}\cdot \mb{x}-\mathcal{E} t)$ &  $0$\\
\hline
 \end{tabular}
\caption{Table of Conserved Charges}\label{table1}
\end{center}
\end{table}
The electric field and magnetic field at $t=0$ is given by,
\begin{eqnarray}
\mb{E}(t=0) &=& \frac{4}{\pi (1+x^2+y^2+z^2)^3} \left[ - 2(xy+z),(x^2-y^2+z^2-1),-2 (yz-x)\right] \nonumber\\
\mb{B}(t=0) &=& \frac{4}{\pi (1+x^2+y^2+z^2)^3} \left[ (x^2-y^2-z^2+1) , 2(xy-z), 2(xz+y) \right] \label{EBt0}
\end{eqnarray}
Then the vector potentials $\mb{A}, \mb{C}$ are given by,
\bea
\mb{C}(t=0) &=& \frac{2}{\pi (1+x^2+y^2+z^2)^2} \left[-z,-1,x\right]\\
\mb{A}(t=0) &=& \frac{2}{\pi (1+x^2+y^2+z^2)^2} \left[1,-z,y\right]
\eea
We can then show that the helicities (in units of energy =1),
\be
h_{ee} =h_{mm} =\half
\ee

The Fourier transform of the Hopfion solution is written in terms of Vector spherical harmonics~\cite{IrvineBouwmeester2008},
\bea
\mb{A}_{lm}^{TE}(k,\mb{r}) &=& \frac{1}{\imath \omega} \frac{j_{l}(k r)}{\sqrt{l (l+1)}} \mb{L} Y_{lm}(\theta, \phi) \\
\mb{A}_{lm}^{TM}(k,\mb{r}) &=& \frac{1}{k^2} \nabla \times \left(\frac{j_{l}(k r)}{\sqrt{l (l+1)}} \mb{L} Y_{lm}(\theta, \phi)\right)
\eea
where $\mb{L}=- \imath \mb{r} \times \nabla$, $Y_{lm}(\theta, \phi)$ are spherical harmonics, $j_{l}(kr)$ are spherical Bessel
functions. Also $l\ge 1$ and $-l < m < l$. Then the vector potential ($B=\nabla \times A$) for the Hopfion is given by~\cite{Ranada:1990} \cite{IrvineBouwmeester2008},
\bea
\mb{A}(\mb{r},t) &=& \int_0^{\infty} dk \left(\mb{A}(k,\mb{r})+\mb{A}(k,\mb{r})^*\right) \\
\mb{A}(k,\mb{r}) &=& \sqrt{\frac{4}{3 \pi}}  k^3 e^{-k} \left[ \mb{A}_{1,1}^{TE}(k,\mb{r}) - \imath \mb{A}_{1,1}^{TM}(k,\mb{r})\right] e^{-\imath \omega t} \label{Hopffourier1}
\eea
The superposition $ \left[ A^{TE} - \imath A^{TM}\right]$ is an eigenstate of curl operator and are known as Chandrasekhar-Kendall
states.
\be
\nabla \times  \left[ \mb{A}_{l,m}^{TE}(k,\mb{r}) \pm \imath \mb{A}_{l,m}^{TM}(k,\mb{r})\right] = \mp k  \left[ \mb{A}_{l,m}^{TE}(k,\mb{r}) \pm \imath \mb{A}_{l,m}^{TM}(k,\mb{r})\right]
\ee
The corresponding electric and magnetic fields are then given by,
\bea
\mb{B}(k,\mb{r}) &=& \nabla \times \left(\mb{A}(k,\mb{r})+\mb{A}(k,\mb{r})^*\right) =  k \left( \mb{A}(k,\mb{r}) + \mb{A}(k,\mb{r})^*\right)\\
\mb{E}(k,\mb{r}) &=& -\partial_t\left(\mb{A}(k,\mb{r})+\mb{A}(k,\mb{r})^*\right)= \imath \omega \left(\mb{A}(k,\mb{r})-\mb{A}(k,\mb{r})^*\right)
\eea
\paragraph{Generalization to $(p,q)$-knots in $(\theta,\phi)$ formulation}
There were various attempts to construct general knotted solution with preserved topological structure in~\cite{Irvine2010},
\cite{Arrayas:2011ia}. But although helicity was conserved, the topological structure varies with time in the class of solution
described in~\cite{Irvine2010},\cite{Arrayas:2011ia}. In a recent paper~\cite{PhysRevLett.111.150404},
authors were able to construct solutions using Bateman's construction~\cite{Bateman} knotted solutions
which preserves its structure over time.
Let us describe the prescription given in ~\cite{Irvine2010}.
We can easily generalize to $(p,q)$-knots by modifying Eq.~(\ref{Hopffourier1}) as given in~\cite{Irvine2010},
\be
\mb{A}(k,\mb{r}) = \sqrt{\frac{4}{3 \pi}}  k^3 e^{-k} \left[ \mb{A}_{1,1}^{TE}(k,\mb{r}) - \imath \frac{p}{q} \mb{A}_{1,1}^{TM}(k,\mb{r})\right] e^{-\imath \omega t} \label{pqfourier1}
\ee
For general $(p,q)$ co-prime integers, this corresponds to all possible Knots. A important point about these solution as
emphasized in~\cite{Irvine2010}, is that $\mb{E} \cdot \mb{B} \ne 0$ but $\int d^3x \mb{E} \cdot \mb{B} =0$. So these cases the helicity is
conserved but if we look at the topological structure of individual field lines, it changes with time. So the knot structure
implied by co-prime integers $(p,q)$ is only valid at $t=0$.

\section{Explicit expression of electric and magnetic field for Hopfion}\label{appendixExplicitEB}

The Riemann-Silberstein vector for the Hopfion solution given by \eqref{hopfalpha1},\eqref{hopfbeta1} is,
\be
\mb{F}=\frac{4}{(A+2 \imath t)^3} \left(\begin{array}{c}
                                   (t-x-z+\imath (y-1)) (t+x-z-\imath(y+1) \\
                                   -\imath (t-y-z-\imath(x+1))(t+y-z+\imath(x-1)) \\
                                   2 (x - \imath y) (t-z-\imath)  \\
                                  \end{array}\right)
\ee
where $ A= ( x^2+y^2+z^2 - t^2 +1)$. The electric and magnetic field can be read of directly from $\mb{F}$
as the real and imaginary part respectively. Generally
the expressions of the electric and magnetic field looks complicated but can be written in a more compact form as
polynomial in $A$. We illustrate that for one of the components for the electric field,
\begin{eqnarray}
E_x &=&  -4 \left [ A^4  -2 A^3 ( y^2+ z^2 - t z) -12 A^2 t( xy +z) +  24 A  t^2( y^2+ z^2  -t z )\right. \nonumber \\
&&\left.+ 16 t^3( -t + xy + z) \right ]/( A^2 + 4 t^2 )^3.
\end{eqnarray}
For $t=0$  this reduces to 
\be
E_x(t=0) = -\frac{ 4}{(1+r^2)^3} ( 1+ x^2 -y^2 -z^2)
\ee 
which is identical to $E_y$ of  (\ref{EBt0}) upon interchanging $x\leftrightarrow y$.
At the origin $\vec r=0$ we find 
\be
E_x = -4\frac{(1-3t^2)}{(1+t^2)^3}
\ee

\end{document}